\documentclass{aastex631}
\newcommand{\RNum}[1]{\uppercase\expandafter{\romannumeral #1\relax}}
\newcommand{\msun}{M_\odot}
\newcommand{\teff}{T_{\rm eff}}

\begin{document}

\title{PRECISE AGE FOR THE BINARY HD 21278 IN THE YOUNG ALPHA PERSEI CLUSTER}

\author{Christopher A. Danner}
\affiliation{San Diego State University, Department of Astronomy, 
San Diego, CA, 92182 USA}

\author[0000-0003-4070-4881]{Eric L. Sandquist}
\affiliation{San Diego State University, Department of Astronomy, 
San Diego, CA, 92182 USA}

\author[0000-0001-5415-9189]{Gail H. Schaefer}
\affiliation{The CHARA Array of Georgia State University, Mount Wilson Observatory, Mount Wilson, CA 91023, USA}

\author[0000-0003-4080-6466]{Luigi R. Bedin}
\affiliation{Istituto Nazionale Astrofisica di Padova—Osservatorio Astronomico di Padova, Vicolo dell'Osservatorio 5, I-35122 Padova, Italy}

\author[0000-0001-9939-2830]{Christopher D. Farrington}
\affiliation{The CHARA Array of Georgia State University, Mount Wilson Observatory, Mount Wilson, CA 91023, USA}

\author[0000-0001-9745-5834]{Cyprien Lanthermann}
\affiliation{The CHARA Array of Georgia State University, Mount Wilson Observatory, Mount Wilson, CA 91023, USA}

\author[0000-0001-6017-8773]{Stefan Kraus}
\affiliation{Astrophysics Group, Department of Physics \& Astronomy, University of Exeter, Stocker Road, Exeter, EX4 4QL, UK}

\author[0000-0002-4313-0169]{Robert Klement}
\affiliation{European Organisation for Astronomical Research in the Southern Hemisphere (ESO), Casilla 19001, Santiago 19, Chile}

\author[0000-0002-2208-6541]{Narsireddy Anugu}
\affiliation{The CHARA Array of Georgia State University, Mount Wilson Observatory, Mount Wilson, CA 91023, USA}

\author[0000-0002-3380-3307]{John D. Monnier}
\affiliation{Astronomy Department, University of Michigan, Ann Arbor, MI 48109, USA}

\author[0000-0001-9647-2886]{Jerome A. Orosz}
\affiliation{San Diego State University, Department of Astronomy, 
San Diego, CA, 92182 USA}

\author[0009-0005-8088-0718]{Isabelle Codron}
\affiliation{Astrophysics Group, Department of Physics \& Astronomy, University of Exeter, Stocker Road, Exeter, EX4 4QL, UK}

\author[0000-0002-3003-3183]{Tyler Gardner}
\affiliation{Cooperative Institute for Research in Environmental Sciences: Boulder, Colorado, US}

\author[0009-0006-0225-4444]{Mayra Gutierrez}
\affiliation{University of California Santa Cruz}

\author[0000-0001-5980-0246]{Benjamin R. Setterholm}
\affiliation{Astronomy Department, University of Michigan, Ann Arbor, MI 48109, USA}

\author[0000-0003-3045-5148]{Jeremy Jones}
\affiliation{The CHARA Array of Georgia State University, Mount Wilson Observatory, Mount Wilson, CA 91023, USA}

\author[0009-0005-8004-2351]{Becky Flores}
\affiliation{The CHARA Array of Georgia State University, Mount Wilson Observatory, Mount Wilson, CA 91023, USA}

\begin{abstract}

We present a study of the double-lined spectroscopic binary HD 21278 that contains one of the brightest main sequence stars in the young $\alpha$ Persei open cluster. We analyzed new spectra and reanalyzed archived spectra to measure precise new radial velocity curves for the binary. We also obtained interferometric data using the CHARA Array at Mount Wilson to measure the sky positions of the two stars and the inclination of the $\sim$ 2 milliarcsecond orbit. We determine that the two stars have masses of $5.381 \pm 0.084 M_{\odot}$ and $3.353 \pm 0.064 M_{\odot}$. From isochrone fits, we find the cluster's age to be $49 \pm 7$ Myr (using PARSEC models) or $49.5 \pm 6$ Myr (MIST models).  Finally, we revisit the massive white dwarfs that are candidate escapees from the $\alpha$ Persei cluster to try to better characterize the massive end of the white dwarf initial-final mass relation. The implied progenitor masses challenge the idea that Chandrasekhar-mass white dwarfs are made by single stars with masses near $8 \msun$. 

\end{abstract}

\section{Introduction} \label{sec:intro}

The age of a stellar system is one of the most difficult characteristics to determine with precision and accuracy.  Mainly, this is because the age of a star is not directly measurable in most cases, requiring comparison to model isochrones.  Because the characteristics of stars on the main sequence generally change very slowly over time, analysis of most main sequence stars will not give precise age estimates. Since the post-main sequence evolution is more rapid, it is precise measurements of stars evolving off of the main sequence that give much more sensitive age estimates. 
This project is the study of one of the most massive, brightest, and evolved stars within the $\alpha$ Persei cluster in order to determine the age of the cluster as a whole.
The star is HD 21278, a double-lined spectroscopic binary with a combined spectral type of B5V  \citep{Zuck12} and a previously-reported orbital period of 21.695 days  \citep{Mor92}.  Based on its photometic and spectroscopic properties, it is likely to
contain one of the brightest main sequence stars in the cluster, nearing the end of its core hydrogen burning life.

$\alpha$ Persei is an open stellar cluster located in the northern part of the Perseus constellation.  According to \cite{GAIA18}, the cluster has a distance of about $175\pm4$ parsecs.  The cluster itself is quite large, spanning over 100 by 200 parsecs \citep{Mei21}. 
\cite{Boy23} present a list of cluster member stars gleaned from a review of eight studies that contained membership lists, including a study by \cite{Lod19} that incorporated membership studies going back to 1956. 

\subsection{The $\alpha$ Per Cluster}

Over the past decades, the metallicity of $\alpha$ Persei has been studied by many groups. \cite{Boy23} cross-matched a list of stars that were rotationally consistent with $\alpha$ Persei members with the LAMOST DR7 LRS Stellar Parameter Catalog \citep{Luo22}. The LAMOST metallicity scale systematically varies with effective temperature \citep{And22}.  They analyzed $\alpha$ Persei's metallicity by comparing its LAMOST effective temperature versus metallicity measurements to that of the Pleiades and Praesepe. They determined that the stellar data from $\alpha$ Persei had a near-overlap of that of the Pleiades, indicating that the two clusters have similar metallicities: marginally super-solar at [Fe/H] $=+0.05 \pm 0.03$.

\citet{cummings} determined a reddening value $E(B-V) = 0.065\pm0.03$ from fits to $UBV$ photometry of vetted members at the cluster turnoff.  \citet{Boy23} derived a reddening value of $E(B-V)=0.058^{+0.032}_{-0.041}$ using STILISM dust maps from \cite{Lal18} and \cite{Cap17}.

Previous age estimates for $\alpha$ Persei all agree that the cluster is very young ($< 100$ Myr), but there is disagreement on a precise value. We summarize some of the more notable studies here. The lithium depletion boundary (LDB) in low-mass stars is generally thought to be an age indicator that is insensitive to stellar physics details, and \citet{galindo} used it to derive an age of $79^{+1.5}_{-2.3}$ Myr. Among isochrone fits to color-magnitude diagram data, \citet{MWSC} derive an age of 50 Myr from 2MASS photometry, and \cite{GAIA18} estimated an age of $71^{+21}_{-18}$ Myr from Gaia photometry. The Gaia fit used PARSEC isochrones, but does not straightforwardly match the brightest evolved cluster stars or the contracting pre-main sequence stars. \cite{Boy23} recently determined $\alpha$ Persei's age relative to the Pleiades and IC 2602 using empirical isochrones based on data from K5V to M3V dwarfs.  They found that $\alpha$ Persei was younger than the Pleiades by $40-50$ Myr and older than IC 2602 by $20-25$ Myr, assuming linear evolution toward the main sequence. Assuming LDB ages for the other two clusters, this resulted in an estimate of $77.5^{+11.9}_{-10.3}$ Myr for $\alpha$ Persei. \citet{Heyl21} determined a "kinematic age" of $81\pm 6$ Myr using candidate pre-main sequence escapees from the cluster --- stars that appear to have left the cluster's vicinity far enough in the past would show significant deviations from the cluster's stellar locus.

One of the best potential applications of an improved age estimate is in the determination of improved initial masses for high-mass white dwarfs.
Three ultramassive white dwarfs were found by \cite{Mil22} and are candidate cluster escapees based on their kinematics.  Each of these white dwarfs has a mass greater than 1 $M_{\odot}$, and one has a mass of 1.2 $M_{\odot}$.

\section{Observations and Methods}

Our primary goal is to precisely measure the masses of the two stars in the HD 21278 binary, and this was accomplished through spectroscopic radial velocity measurements and interferometric measurements of the relative sky positions of the two stars.

\subsection{Spectroscopy\label{specsec}}

To measure radial velocities, we utilized a total of 31 spectra from four different telescopes.  \cite{Mor92} obtained 14 spectra of the system using the 1 meter coud\'{e} feed telescope and spectrograph at Kitt Peak National Observatory (KPNO; \cite{Mor92}). We were able to obtain eleven of these spectra in electronic form (N. Morrell, private communication).  Each spectrum was taken at a different phase ($\phi$) in the binary's orbit. The spectrograph had a resolution of 0.22 \AA \: or 15 km s$^{-1}$ per pixel. Each spectrum covered the range $4320-4500$ \AA \: with the He I 4471 \AA \: and Mg II 4481 \AA \: absorption lines included. We redid the continuum normalization of the spectra outside of obvious absorption lines using a polynomial fit. The radial velocity measurements are described in more detail below.

We obtained eleven spectra in 2020 using the 2.6 m Nordic Optical Telescope (NOT) at the Roque de los Muchachos Observatory.  The spectra were taken using the FIbre-fed Echelle Spectrograph (FIES; \citealt{Tel14}), which has a spectral resolution of 67,000.  The spectra had a wavelength range of $3700-8300$ \AA, from the optical to the infrared. Our analysis used the $4000-6500$ \AA \: range, including He I 4471 \AA,  Mg II 4481 \AA, and Si II 6347 \AA \: and 6371 \AA \: absorption lines, but excluding the strong Balmer lines.
We masked out the Balmer lines because they appeared to introduce a bias on the radial velocities, probably related to the large collisional broadening.

Finally, we obtained archival spectra from two other sources. We collected spectra taken with the Narval spectropolarimeter on the Bernard Lyot Telescope (TBL) from the PolarBase archive \citep{polarbase}. Narval spectra have high spectral resolution ($R \sim 68000$), and cover the wavelength range from about $3750-10500$ \AA  with 40 echelle orders. For HD 21278, 8 intensity spectra were taken (PI: C. Neiner) split between two successive nights in September 2013. We continuum-normalized the spectra, and combined echelle orders using the Pyraf routine {\tt scombine}. For the measurement of radial velocities, we masked the spectra to the same wavelength range as the FIES spectra. These spectra have the highest signal-to-noise, and there is very good agreement between the velocities taken within a night, indicating a precision of $0.1-0.4$ km s$^{-1}$. Because of this, we averaged the measurements made on each night. Using Narval spectra, we measure the rotational velocity of the primary to be $53.4\pm0.3$ km s$^{-1}$ and the secondary to be $9.3\pm0.1$ km s$^{-1}$.

We also obtained one spectrum from the Mercator Library of High-Resolution Stellar Spectroscopy (MELCHIORS; \citealt{melchior}), taken with the Mercator-HERMES spectrograph ($R = 85000$) at the Roque de los Muchachos observatory in September 2011. We used the telluric-corrected and normalized spectrum from the library.

We measured radial velocities from broadening functions (BFs) determined for each spectrum using the program {\tt BF\_python}\footnote{https://github.com/mrawls/BF-rvplotter}, based on algorithms by \cite{Ruc92}. 
The observed spectrum $P(\lambda)$ is assumed  to be a convolution of a sharp-lined spectrum $S(\lambda)$ (a theoretical spectrum is usually used in practice) with the broadening function $B(\lambda)$ \citep{Ruc99}.  The width of the BF reflects the rotational broadening, while the peak gives the radial velocity of the star.  Additionally, the area under each BF peak can be useful for determining the luminosity ratios if the two stars share roughly the same temperature. 
We modified {\tt BF\_python} to fit the BF with either Gaussians or rotational broadening functions. 

For the sharp-lined spectrum, we used synthetic spectra from the high-resolution library BLUERED \citep{Ber08}.  After some experimentation, we decided to use a spectrum that had a temperature of $14000$ K and a surface gravity $\log_{10} g = 4$ (cgs units), as it maximized the signal from both stars.

\begin{figure}[h] %Trying h instead of ht
  \centering
  \begin{minipage}{6.5in}
    \includegraphics[width=\linewidth]{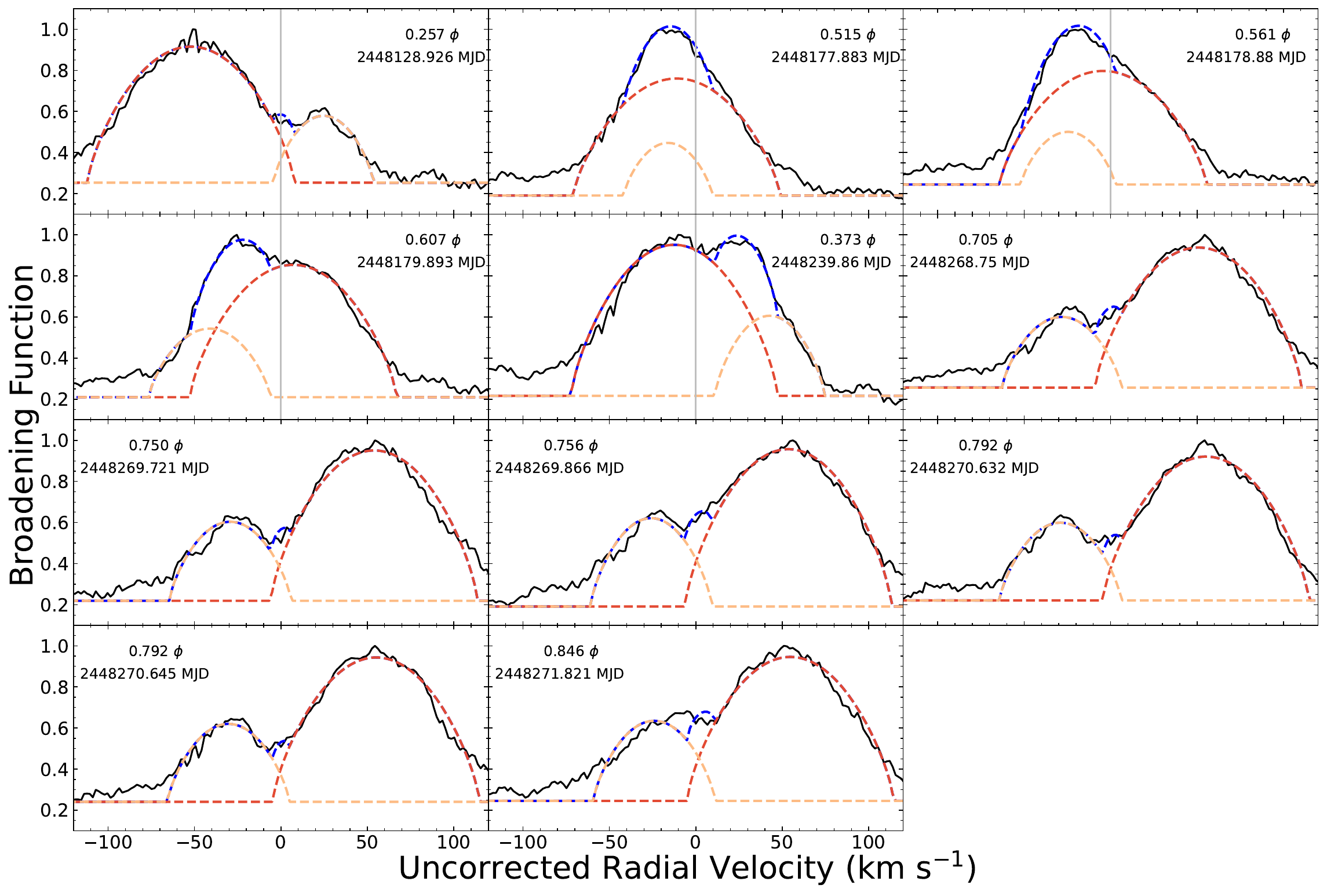}
    \caption{Broadening functions for KPNO spectra.  The red dashed line is the rotational fit for the primary star, the gold dashed line is the rotational fit for the secondary star, the blue dashed line is the combined fit, and the black line is the measured broadening function.  \label{fig2.1}}
  \end{minipage}
\end{figure}

Because of the relatively low resolution of the original KPNO observations and the untargeted timing of the observations, the BF peaks of the two stars overlapped in many spectra (see Fig. \ref{fig2.1}). During our analysis, we found degeneracies in the fits between the measured BF widths and the radial velocities when the secondary star peak overlapped the wing of the primary star peak.  To minimze this, we determined the rotational broadening for each star from the spectra with the best velocity separation of the peaks, and then fit the BFs with more restricted ranges for the rotational velocities. Because the BF peak for the faster-rotating primary star appears largely the same in the different spectroscopic data sets, our initial guesses on the width and amplitudes of the primary star peak could be derived from higher resolution spectra.  

\begin{figure}[ht]
  \centering
  \begin{minipage}{6.5in}
    \includegraphics[width=\linewidth]{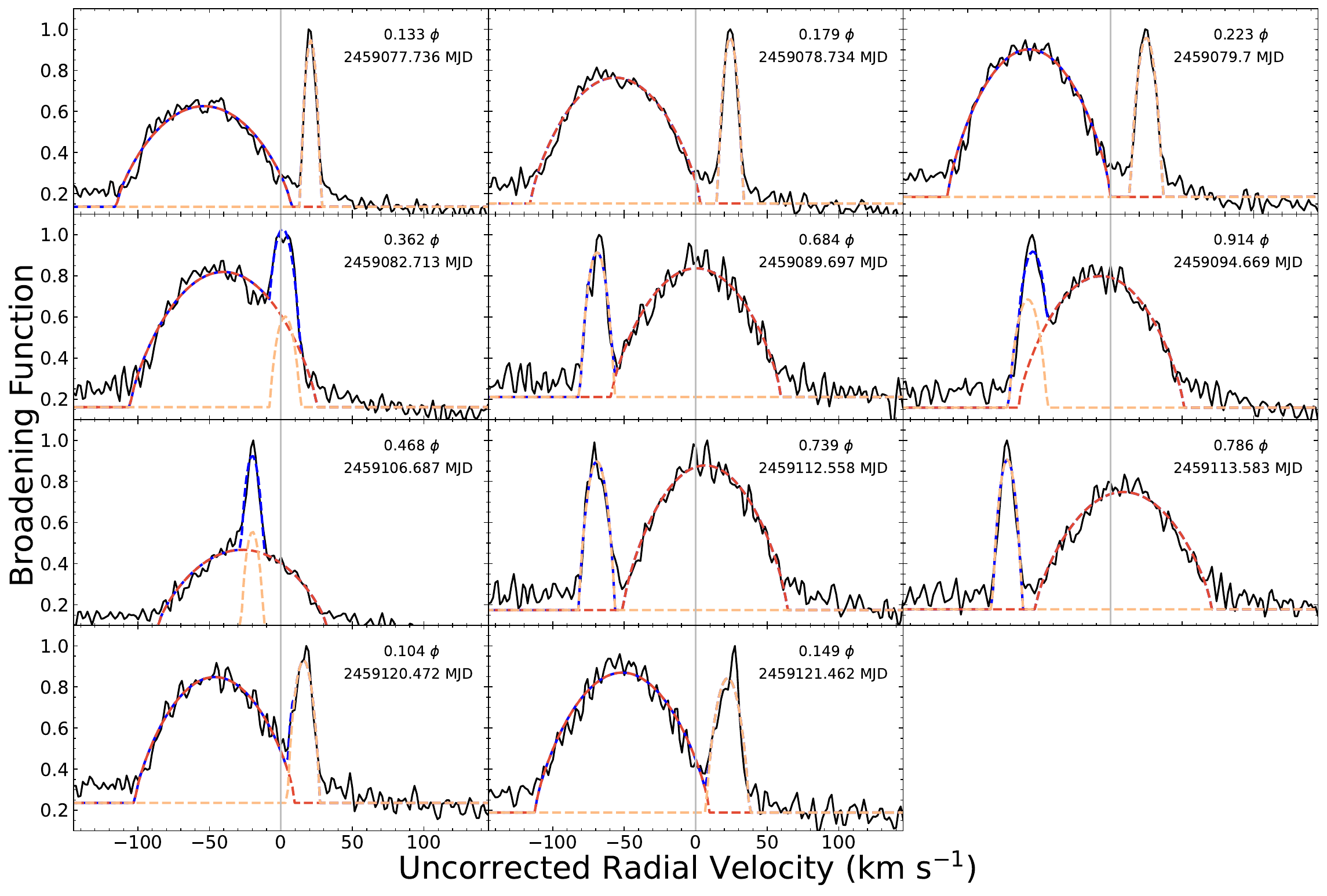}
    \caption{Broadening functions for the FIES spectra.  The line meanings are the same as Fig. ~\ref{fig2.1}. 
  \label{fig2.2}}
  \end{minipage}
\end{figure}

Because we knew the orbital ephemeris from the \cite{Mor92} study, we were able to time our FIES observations to phases when the stars were at their maximum radial velocity separation.   The spectrograph also had higher resolution than the KPNO coud\'{e}, giving much more distinguishable peaks for both stars.  As demonstrated in Fig.~\ref{fig2.2}, the secondary star peak is very sharp. As can be seen in the lower panels of Fig. ~\ref{fig3.1}, the fit residuals for both the primary and secondary star are consistent for the KPNO and NOT velocities at phases where they overlap, giving us confidence that line blending did not produce systematic effects on the velocities. 

From the fits, we extracted 25 radial velocity measurements for each star, and we included three primary star measurements from \cite{Mor92} for phases near zero crossings, when the secondary star would have affected the measurements minimally. The radial velocity measurements for each star are provided in Table \ref{rvtab}. The reported uncertainties are scaled from the uncertainties returned by the {\tt BF\_python} code.  Specifically, we fitted the KPNO and NOT radial velocity datasets separately and scaled the uncertainties by a factor needed to produce a reduced $\chi^2$ value of 1. Because we will ultimately fit a merged dataset of velocities and interferometric observables, we need to have a realistic idea of how each dataset should be weighted, as each influences the orbital parameters in somewhat different directions.  The secondary star BF peak in one spectrum had a fairly significant distortion, and we decided to increase the uncertainty to 3.0 km s$^{-1}$.

\begin{deluxetable}{lrrrrl}[ht]
\tabletypesize{\footnotesize}
\tablewidth{0pt}
\tablecaption{Radial Velocity Measurements \label{rvtab}}
\tablehead{\colhead{HJD$-2400000$} & \colhead{$v_1$}  & \colhead{$\sigma_{v_1}$} & \colhead{$v_2$} &\colhead{$\sigma_{v_2}$} & \colhead{Instrument} \\ & \multicolumn{2}{c}{(km s$^{-1}$)} & \multicolumn{2}{c}{(km s$^{-1}$)} & }
\startdata 
47581.741 & $-2.19$ & 3.00 & & & KPNO/coud\'{e} feed \\
47852.821 & 0.01 & 3.00 & & & KPNO/coud\'{e} feed \\
47896.718 & 2.01 & 3.00 & & & KPNO/coud\'{e} feed \\
48128.926 & $-29.30$ & 1.79 & 48.00 & 2.44 & KPNO/coud\'{e} feed \\ 
48177.883 & 3.10 & 2.01 & $-1.60$ & 3.56 & KPNO/coud\'{e} feed \\ 
48178.880 & 10.26 & 2.01 & $-11.44$ & 3.95 & KPNO/coud\'{e} feed \\ 
48179.893 & 22.37 & 2.27 & $-28.15$ & 2.33 & KPNO/coud\'{e} feed \\ 
48239.860 & $-25.27$ & 1.83 & 30.57 & 2.21 & KPNO/coud\'{e} feed \\ 
48268.750 & 28.07 & 1.75 & $-46.95$ & 1.97 & KPNO/coud\'{e} feed \\ 
48269.721 & 30.35 & 1.70 & $-49.21$ & 2.00 & KPNO/coud\'{e} feed \\ 
48269.866 & 31.03 & 1.46 & $-46.85$ &  1.94 & KPNO/coud\'{e} feed \\ 
48270.632 & 30.02 & 1.58 & $-53.62$ & 2.22 & KPNO/coud\'{e} feed \\ 
48270.645 & 30.65 & 1.22 & $-52.38$ & 2.15 & KPNO/coud\'{e} feed \\ 
48271.821 & 30.82 & 1.69 & $-44.99$ & 2.13 & KPNO/coud\'{e} feed \\ 
56558.59979 & 6.34 & 0.04 & $-10.30$ & 0.10 & TBL/Narval \\
56559.60725 & $-3.63$ & 0.41 & 10.48 & 0.30 & TBL/Narval \\
55809.755079 & $-8.00$ & 0.14 & 13.75 & 0.06 & Mercator/HERMES \\
59077.736 & $-28.33$ & 1.23 & 46.28 & 0.67 & NOT/FIES \\ 
59078.734 & $-30.61$ & 1.07 & 49.68 & 0.76 & NOT/FIES \\ 
59079.700 & $-31.23$ & 0.90 & 50.71 & 0.86 & NOT/FIES \\ 
59082.713 & $-14.37$ & 1.07 & 28.73 & 1.62 & NOT/FIES \\ 
59089.697 & 26.08 & 1.08 & $-43.14$ & 1.01 & NOT/FIES \\ 
59094.669 & 18.87 & 1.28 & $-32.23$ & 1.27 & NOT/FIES \\ 
59106.687 & $-2.62$ & 1.65 & 4.46 & 1.20 & NOT/FIES \\ 
59112.558 & 30.09 & 0.89 & $-45.59$ & 1.10 & NOT/FIES \\ 
59113.583 & 32.33 & 1.16 & $-48.91$ & 0.87 & NOT/FIES \\ 
59120.472 & $-24.27$ & 0.85 & 37.76 & 1.05 & NOT/FIES \\ 
59121.462 & $-29.71$ & 1.01 & 43.98 & 3.00 & NOT/FIES \\ 
\enddata
%\vspace{-0.5cm}
\end{deluxetable}

While there are other archival radial velocity measurements in the literature \citep{frost,petrie,abtmwo,kodaira,abtlevy,gomezabt,fehrenbach,morse},
these were most often measurements of one star. As we have seen from the broadening functions, there is never an orbital phase when the two stars have completely separated lines, and the reality is that velocity measurements are biased when the lines blend so strongly. We found systematic deviations from the radial velocity curve determined from the spectra, and therefore did not use the literature velocities further.

\begin{figure}[h] %Trying h instead of ht
  \centering
  \begin{minipage}{6.5in}
    \includegraphics[width=\linewidth]{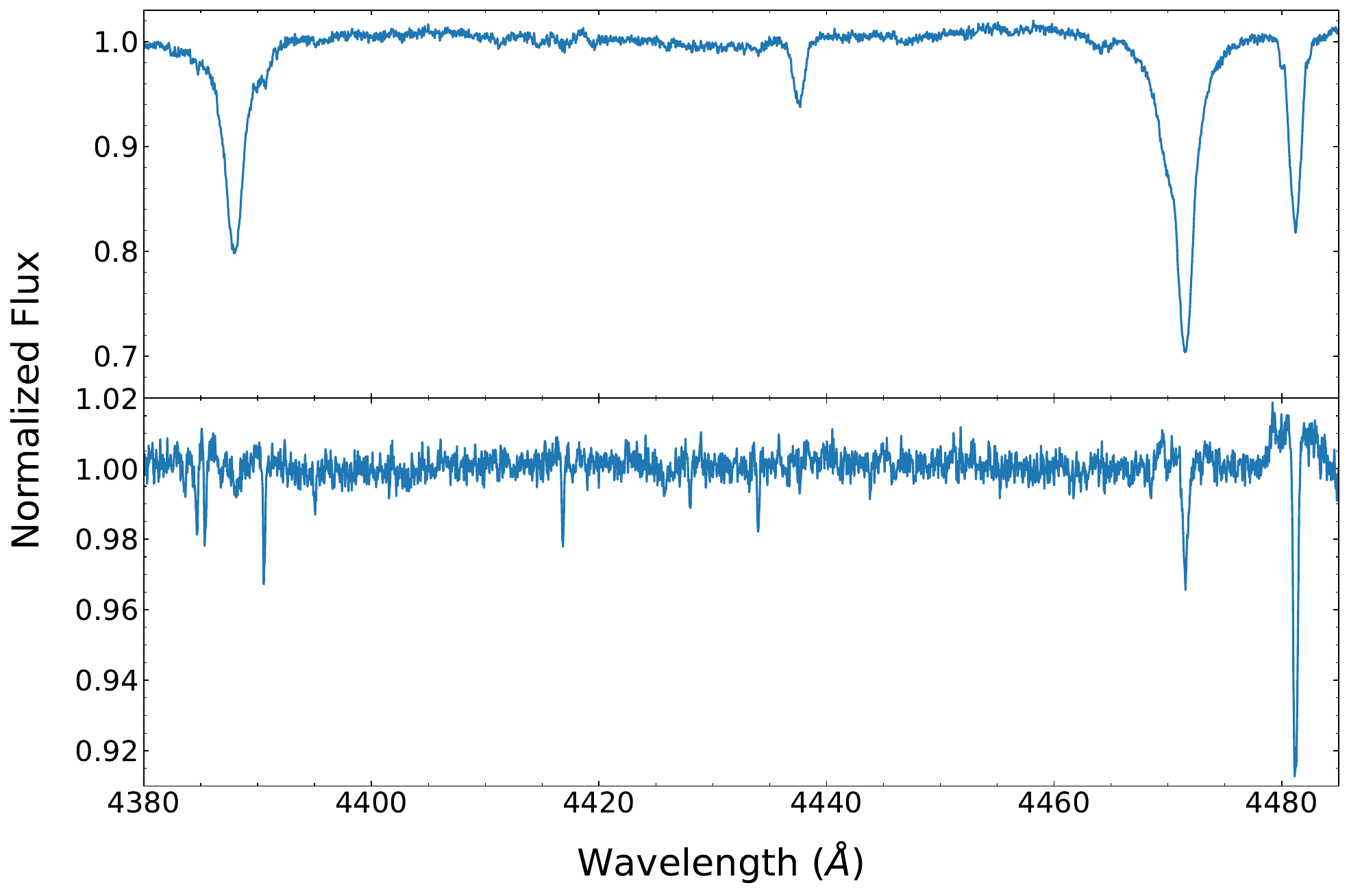}
    \caption{Disentangled spectra of the HD 21278 binary for the primary star (top panel) and secondary star (bottom panel).  The He I 4387 and 4471 \AA \: absorption lines, and the Mg II 4481 \AA \: absorption line are the most prominent features. 
  \label{fig2.3}}
  \end{minipage}
\end{figure}

To more closely examine the spectra for each star, we used a disentangling algorithm \citep{Gon06}, which employs spectra at different orbital phases to iteratively determine radial velocities for each spectrum and averaged spectra for each star.  We compared the radial velocity outputs from disentangling and BF fitting, and found good agreement. While it is possible to use distinct synthetic spectra for each star to initiate the disentangling process, we obtained satisfactory results using the same 14000 K synthetic spectrum for both stars.  The disentangled average spectra are shown in Fig. \ref{fig2.3}.

With the disentangled spectra, we determined the temperatures and spectral types of the two stars by measuring the He\RNum{1} $\lambda 4471$/ Mg\RNum{2} $\lambda 4481$ line ratios of both stars.  Previous studies \citep{Ram20} provided a rough classification for B-type stars based on the line ratios of  He\RNum{1} $\lambda 4471$ and Mg\RNum{2} $\lambda 4481$.  According to them, if He\RNum{1} $\lambda 4471$ is greater than Mg\RNum{2} $\lambda 4481$, then the star has a spectral type around B3-B5 and if He\RNum{1} $\lambda 4471$ is less than Mg\RNum{2} $\lambda 4481$, then the star has a spectral type around B8-A0.

We estimated the temperatures of the two stars by comparing the He\RNum{1} $\lambda 4471$/ Mg\RNum{2} $\lambda 4481$ line ratios of both stars to ATLAS9 synthetic spectra \citep{Ber08}.   
Each synthetic spectrum had solar metallicity and a surface gravity $\log g = 4.0$.  
A rough estimation of the temperatures for the primary and secondary stars were determined from interpolation in the line ratio array: approximately $16,750$ K for the primary and approximately $11,120$ K for the secondary.  This indicates that the primary star is most likely a B5 star while the secondary star is likely a B9 star, which lines up nicely with the line ratio to B-type classification from \cite{Ram20}.

\subsection{Interferometry}
We obtained interferometric data for HD 21278 to resolve the sky-projected orbit of the binary, and to obtain luminosity ratio information for the component stars.  Our observations were taken at the Center for High Angular Resolution Astronomy (CHARA; \cite{Bru16}) Array  at Mt. Wilson, composed of six 1-meter telescopes with a maximum baseline of 331 meters between any pair of telescopes.  Over the course of this project, the CLIMB \citep{Bru05}, MIRC-X \citep{Ang20}, and MYSTIC \citep{Set23} beam combiners  were used for HD 21278. We observed the binary on  two nights in 2018 with the CLIMB three-telescope beam combiner, and on six nights using MIRC-X in {\it H} band and MYSTIC in the {\it K} band in 2021, 2022, and 2023.  The MYSTIC and MIRC-X data were taken in the low resolution Prism 50/ Prism 49 modes, respectively.  
The MIRC-X beam combiner had recently been upgraded to improve sensitivity and wavelength coverage.
Both MYSTIC and MIRC-X are very well suited for detecting binary companions because the simultaneous usage of all six telescopes in the array allowed for observations on many different baselines.

In addition to the data collected at CHARA, we also utilized two sets of interferometric data that were taken at the Palomar Testbed Interferometer (PTI; \cite{Col199}, \cite{Col299}) in late 2002. PTI is composed of three 40 cm telescopes (although only two were used at a time) with a maximum baseline of 110 meters.  The PTI data were taken in the {\it K} band. 

Interferometry can produce position information and flux ratio information from the fringe visibility and closure phase.  The fringe visibilities of individual stars are complex quantities, and $\mathcal{V}^2_{obs}$ is used as the observable.  For a binary system, this is modeled as
\begin{equation} \label{eq:1.10}
    \mathcal{V}_{binary}^2 = \frac{V_1^2 + r^2V_2^2+2r|V_1||V_2|\cos[2\pi (u\Delta \alpha + v \Delta \beta)]}{(1+r)^2} ,
\end{equation}
where $V_1$ and $V_2$ are the complex visibilities of the two stars, $r$ is the ratio of the power ($P_2/P_1$) of the two stars, $\Delta \alpha$ and $\Delta \beta$ are the angular offsets of the secondary star relative to the primary, and $u$ and $v$ are the angular spatial frequencies \citep{Tra00}. 

Wavefronts from the star reach the telescopes in the array at different times, which results in a geometric path-length difference that is compensated for with delay-line carts.  Atmospheric refraction causes additional path-delay differences between the telescopes that are canceled out by computing the closure phase between telescope triplets in the array.
Aside from providing information regarding the asymmetry of the light distribution, closure phases can give a precise measurement of the flux ratio for stars in a binary system.

The CLIMB data were reduced using the pipeline developed by J. D. Monnier, with the general method described by \cite{Mon11} and extended to three beams (e.g., \cite{Klu18}), producing broadband squared visibilities for each baseline and closure phases for each closed triangle.
The MIRC-X and MYSTIC data were reduced using the standard pipeline\footnote{ \url{https://gitlab.chara.gsu.edu/lebouquj/mircx_pipeline.git}} (versions 1.3.5 and 1.4.0; \citealt{Angu20}) that produces squared visibilities and triple products for each baseline and spectral channel.  For the MIRC-X and MYSTIC observations, a 5\% uncertainty on visibility measurements and a $0\fdg3$ uncertainty on closure phases were added in quadrature 
with statistical uncertainties to account for uncertainties in calibration. Systematic errors in wavelength can lead to systematics in the angular scale, which potentially affects the measurement of the orbit size. A small systematic wavelength offset  was identified \citep[e.g.,][]{gardner21} and applied (wavelengths divided by a factor of 1.0054 for MIRC-X and 1.006 for MYSTIC; J. D. Monnier, private communication), and a small uncertainty due to wavelength calibration uncertainty (0.2\%) was added to the angular distance measurements.

As we did with the radial velocity measurements, we scaled the measurement uncertainties to force the reduced $\chi^2$ to 1 for the best fit to the interferometric data. The data were scaled separately for each beam combiner.  The interferometric data are summarized in Table \ref{tab1}. 

\begin{deluxetable}{lllllrrl}

\tabletypesize{\footnotesize}
\tablewidth{0pt}

\tablecaption{Interferometer Data \label{tab1}}

\tablehead{\colhead{Facility} & \colhead{Combiner} & \colhead{Telescopes} & \colhead{UT Date} & \colhead{Time}& \colhead{$N_V$} & \colhead{$N_{CP}$} & \colhead{Calibrators}}

\startdata 
PTI & NW & 3 & 9/24/2002 & 52541.9362 & 3 & 0 & -- \\ 
PTI & NW & 3 & 12/12/2002 & 52620.75442 & 1 & 0 & -- \\ 
CHARA & CLIMB & S1,W1,E1 & 11/18/2018 & 58441.73946 & 35 & 5 & HD 21363 \\ 
CHARA & CLIMB & S1,W1,E1 & 11/25/2018 & 58447.92866 & 14 & 2 & HD 21363 \\ 
CHARA & MIRC-X & E1,W2,W1,S2,S1,E2 & 11/18/2021 & 59536.371 & 432 & 496 & HD 21268, HD 21363 \\ 
CHARA & MYSTIC & E1,W2,W1,S2,S1,E2 & 11/18/2021 & 59536.371 & 660 & 880 & HD 21268, HD 21363 \\ 
CHARA & MIRC-X & E1,W2,W1,S2,E2 & 12/11/2021 & 59559.368 & 320 & 320 & HD 21268 \\ 
CHARA & MYSTIC & E1,W2,W1,S2,E2 & 12/11/2021 & 59559.368 & 403 & 400 & HD 21268 \\ 
CHARA & MIRC-X & E1,W2,W1,S2,E2 & 12/19/2021 & 59567.368 & 320 & 320 & HD 21085 \\ 
CHARA & MYSTIC & E1,W2,W1,S2,E2 & 12/19/2021 & 59567.368 & 440 & 440 & HD 21085 \\  
CHARA & MIRC-X & E1,W2,W1,S1,E2 & 11/14/2022 & 59897.372 & 280 & 280 & HD 21363 \\ 
CHARA & MYSTIC & E1,W2,W1,S1,E2 & 11/14/2022 & 59897.372 & 440 & 440 & HD 21363 \\ 
CHARA & MIRC-X & E1,W2,W1,S1,E2 & 11/21/2022 & 59904.362 & 320 & 320 & HD 21085, HD 21363 \\ 
CHARA & MYSTIC & E1,W2,W1,E2 & 11/21/2022 & 59904.362 & 264 & 176 & HD 21085, HD 21363 \\ 
CHARA & MIRC-X & E1,W2,W1,S2,S1,E2 & 9/28/2023 & 60215.314 & 480 & 640 & HD 21363 \\
CHARA & MYSTIC & E1,W2,W1,S2,S1,E2 & 9/28/2023 & 60215.314 & 580 & 760 & HD 21363 \\
\enddata

%\vspace{-0.5cm}
\tablecomments{Time = HJD $- 2,400,000$}

\end{deluxetable}

\section{Binary Modeling}
%\subsection{Inputs}

We model HD 21278 with a two-body Keplerian orbit, and fit radial velocity measurements for each star, interferometric visibilities, and interferometric closure phases using a program called {\tt interfRVorbit}. (Radial velocity measurements from Narval and Melchior spectra were not included in the final fit because both sets taken near velocity crossings and their low measurement uncertainties were noticeably pulling the solution off the best fit to the velocity extremes.)
The interfRVorbit program uses a genetic algorithm \citep{Cha95} to fit the data, and returns best-fitting orbital parameters for the binary system. 
We fit thirteen parameters and set the ranges to be searched.  
The velocity semi-amplitudes $K_1$ and $K_2$, the systemic velocity $\gamma$, eccentricity $e$, orbital period $P$, argument of periastron $\omega$, and time of periastron $t_0$ are all parameters that can be constrained from a fit to only the radial velocities.  
The remaining parameters (inclination angle $i$, angle of ascending node $\Omega$, binary angular size $a''$, and the flux ratios $L_2/L_1$ of the two stars in different wavelength bands) can only be determined from fits that utilize relative sky position information from the interferometry.   As demonstrated in Table \ref{tab2}, these parameters are quite precise, primarily due to the large numbers of fringe visibility and closure phase constraints on the relative orbit. By way of comparison, \citep{Tor24} conducted a similar analysis on the binary HD 174881 using a factor of 10 fewer visibility measurements and few closure phases, and found uncertainties for parameters such as $\Omega$ and $i$ about $3-5$ times larger compared than ours.

\begin{figure}[ht]
  \centering
  \begin{minipage}{6in}
    \includegraphics[width=\linewidth]{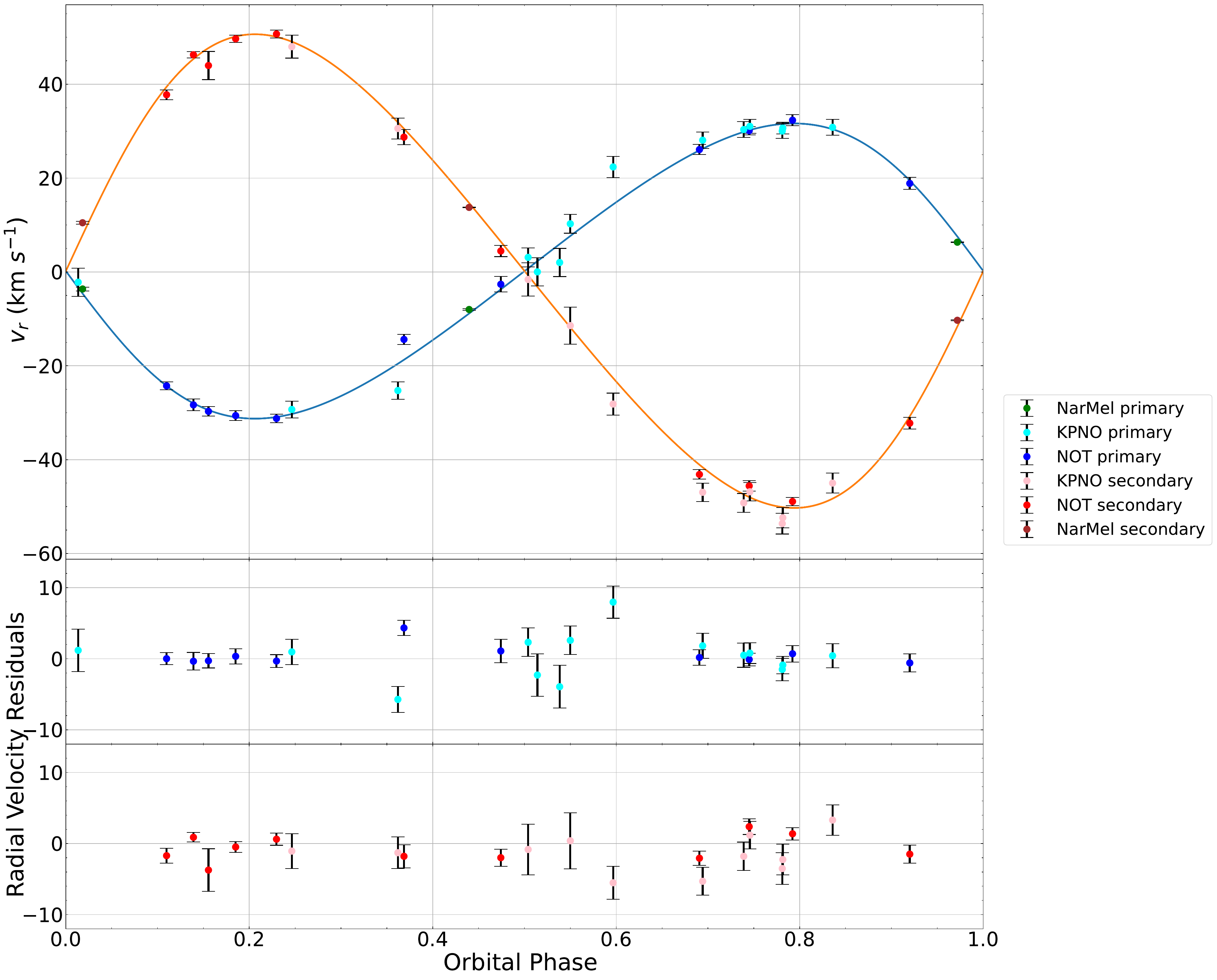}
    \caption{Top: Radial velocity measurements from KPNO (cyan and pink) and NOT (blue and red) spectra versus orbital phase, and best-fit curve. Radial velocity measurements from Narval \citep{polarbase} and Melchior \citep{melchior} are included (green and brown) in the best-fit curve. Middle/Bottom: Velocity residuals (observed minus model) versus orbital phase for the primary (middle) and secondary (bottom). 
  \label{fig3.1}}
  \end{minipage}
\end{figure}

We calculated the parameter uncertainties by finding the ranges covered by models with $\chi^2$ within 1
of the minimum value \citep{Avni76}.  A $\Delta \chi^2$ value of 1 for an individual parameter roughly corresponds to the parameter's $68\%$ confidence level. We used the best fit velocity semi-amplitudes, inclination, eccentricity, and period to estimate the masses of the primary and secondary star.
From the overall fit, we found that the primary star has a mass of $5.381 \pm 0.084 M_{\odot}$ and the secondary star has a mass of $3.353 \pm 0.064 M_{\odot}$. Using the same best-fit parameters that we had used to estimate the individual masses of the two stars, we also estimated the semi-major axis separation at $0.3134\pm0.0017$ AU.  The semi-major axis separation was then used in tandem with the period to estimate that the total mass of the system is $8.73\pm0.14 M_{\odot}$.  We made a separate a total mass estimate using only the period, the angular size, and the Gaia distance of the binary for comparison.  Using only those three variables, the total mass was $8.28\pm0.56 M_{\odot}$.  The total mass uncertainty was much higher compared to the other because the Gaia distance has a 2.3 percent distance uncertainty that is amplified in the total mass calculation.  This is preventing the interferometric total mass from being as precise as the other one.  Using the semi-major axis separation and the best-fit angular size of the semi-major axis, we estimated that the binary has a distance of $178.26\pm0.96$ pc.  This falls within the range of uncertainty of the $175.11\pm3.96$ pc distance given by \cite{Bai21}.

\begin{deluxetable}{llll}
\tabletypesize{\footnotesize}
\tablewidth{0pt}
\tablecaption{Best-Fit Stellar Parameters \label{tab2}}
\tablehead{\colhead{Parameter} & \colhead{M\&A92} & \colhead{RV-Only Fit} & \colhead{All Parameter Fit}}
\startdata 
$K_1$ (km s$^{-1}$) & 22.7 $\pm$ 0.9 & 31.35 $\pm$ 0.32 & 31.45 $\pm$ 0.31 \\ 
$K_2$ (km s$^{-1}$) & 49.0 $\pm$ 3.0 & 50.29 $\pm$ 0.33 & 50.47 $\pm$ 0.31 \\ 
$e$ & 0.12 $\pm$ 0.04 & 0.1282 $\pm$ 0.0074 & 0.13843 $\pm$ 0.00010 \\ 
$\omega$ $(^\circ)$ & 109 $\pm$ 3 & 95.3 $\pm$ 2.9 & 89.946 $\pm$ 0.049 \\ 
$\gamma$ (km/s) & -- & 0.13 $\pm$ 0.20 & 0.19 $\pm$ 0.19 \\ 
$P$ (days) & 21.695 $\pm$ 0.004 & 21.68564 $\pm$ 0.00017 & 21.685415 $\pm$ 0.000035 \\ 
$t_0$ (mJD) & 46714.5 $\pm$ 0.2 & 46714.152 $\pm$ 0.202 & 46714.031 $\pm$ 0.022 \\ 
$\Omega (^\circ)$ & -- & -- & 85.725 $\pm$ 0.049 \\ 
$i (^\circ)$ & -- & -- & 148.938 $\pm$ 0.030 \\ 
$a''$ (mas) & -- & -- & 1.75820 $\pm$ 0.00377 \\ 
$L_2/L_1$ ($H$) & -- & -- & 0.25047 $\pm$ 0.00019 \\ 
$L_2/L_1$ ($K$) & -- & -- & 0.25558 $\pm$ 0.00020 \\ 
%$a$ (mas) & -- & -- & 0.2500 $\pm$ 0.0015 \\ 
$M_1$ ($M_{\odot}$) & -- & -- & 5.381 $\pm$ 0.084\\ 
$M_2$ ($M_{\odot}$) & -- & -- & 3.353 $\pm$ 0.064 \\
$M_1+M_2$ ($M_{\odot}$) & -- & -- & 8.735 $\pm$ 0.142 \\
\enddata
%\vspace{-0.5cm}
\end{deluxetable}

\begin{figure}[ht]
  \centering
  \begin{minipage}{6in}
    \includegraphics[width=\linewidth]{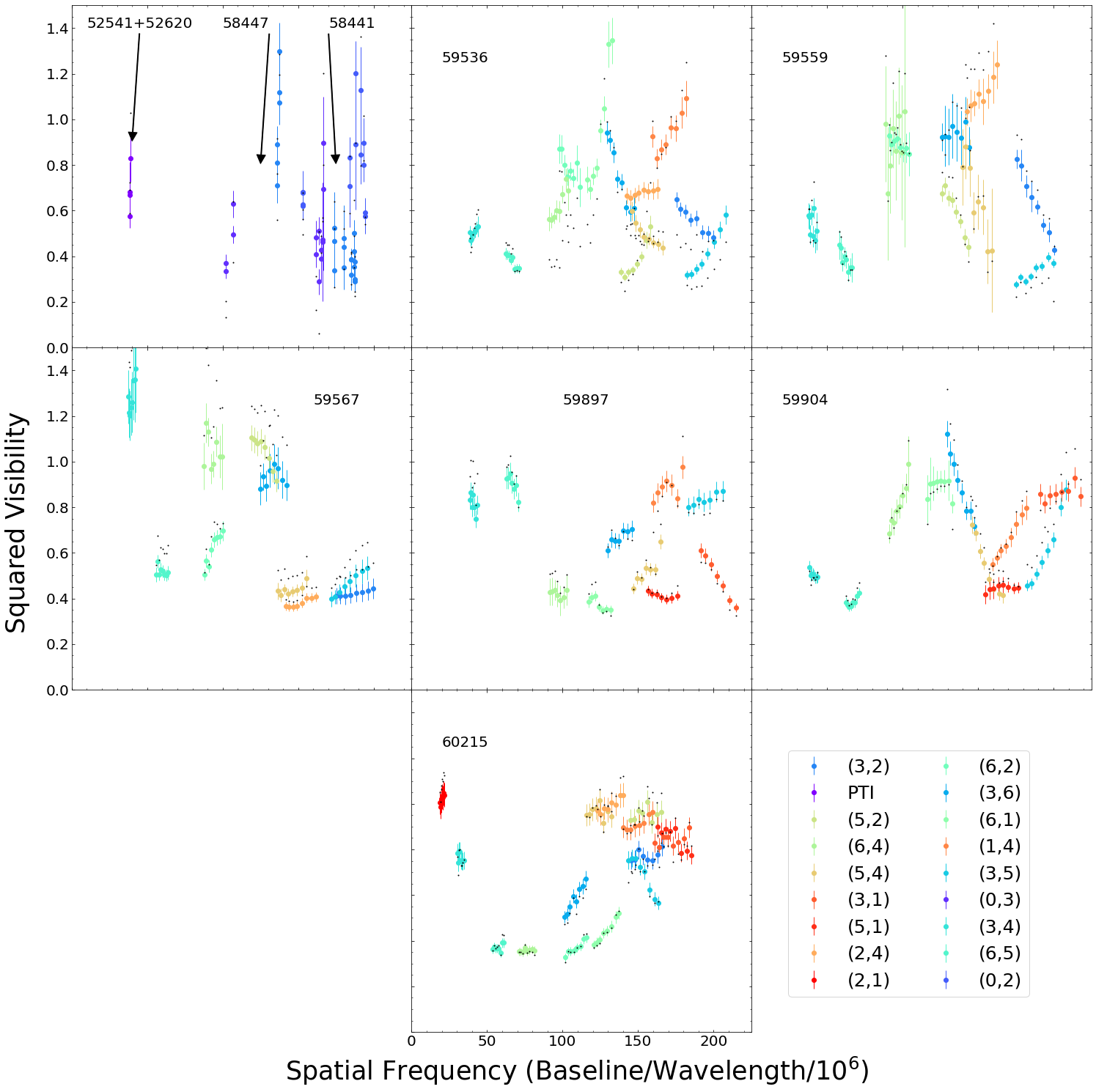}
    \caption{Squared visibilities versus spatial frequency for MIRC-X observations and predictions of the best-fit model (black dots), separated by  mJD of observation. For clarity, only observations taken at one epoch are shown for each night.  Observations are separated by telescope pair with each pair denoting the two telescopes that were used to obtain visibility measurements. The top left panel has four epochs of PTI and CLIMB observations. 
  \label{fig3.2}}
  \end{minipage}
\end{figure}

\begin{figure}[ht]
  \centering
  \begin{minipage}{6in}
    \includegraphics[width=\linewidth]{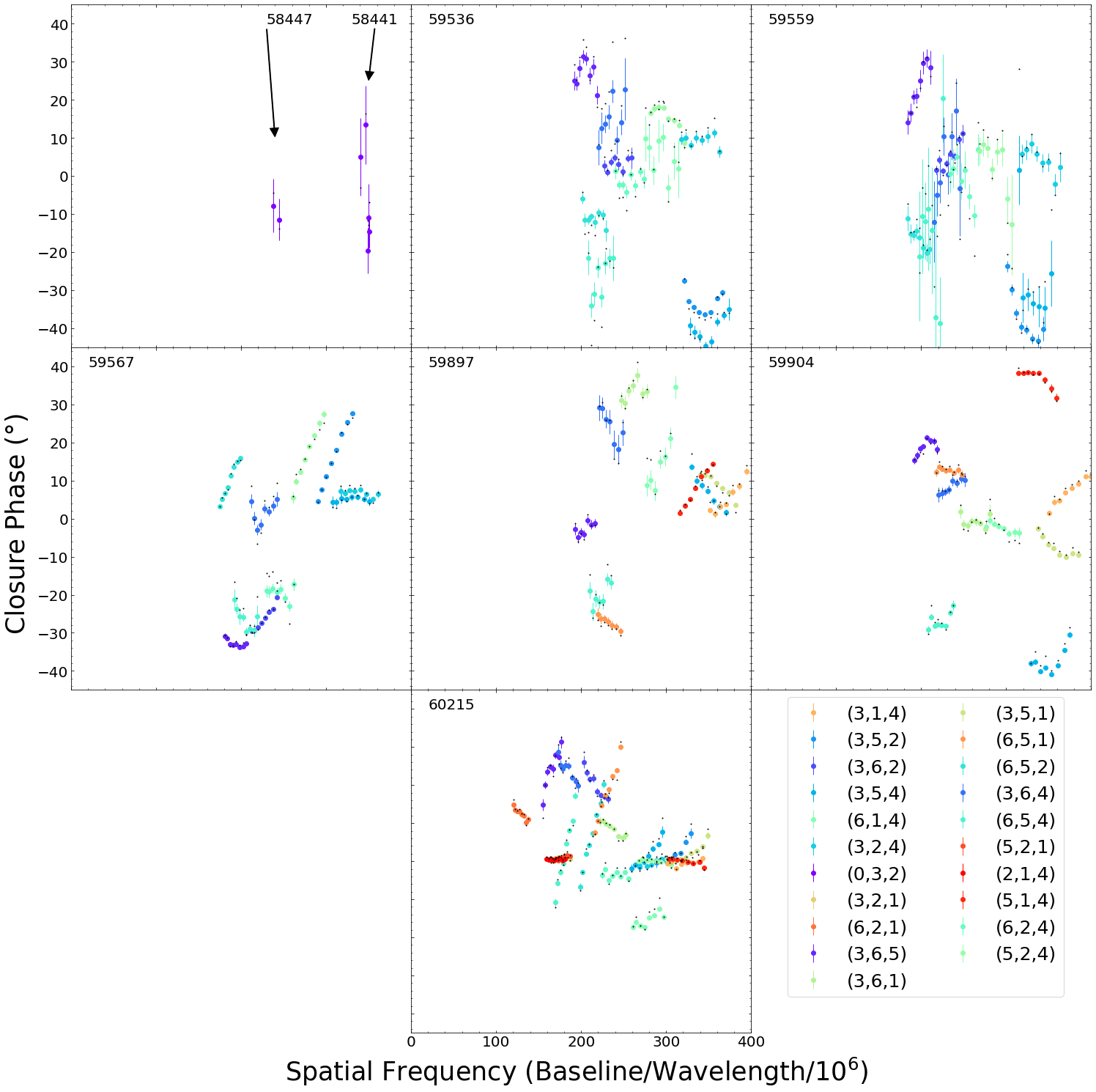}
    \caption{Closure phase versus spatial frequency for MIRC-X observations and predictions of the best-fit model (black dots), separated by mJD of observation. The plotted spatial frequency is obtained by summing for two of the three baselines.  For clarity, only observations taken at one epoch are shown for each night.  The top left panel has two epochs of CLIMB observations. Each closure phase point utilizes 3 different telescope apertures, as listed in the legend.  
  \label{fig3.3}}
  \end{minipage}
\end{figure}

We compare the model to observed radial velocities in Fig. \ref{fig3.1}, and
compare the model and observed visibilities and closure phases versus the baseline in Figs.~\ref{fig3.2} and \ref{fig3.3}. Most of the visibility and closure phase predictions appear to agree with the data points with two notable exceptions.  On mJDs 59536 and 59559, a sizable portion of both the visibility and closure phase data points do not match with the corresponding model points.  This may be connected to poor seeing conditions and high humidity on those nights.

\begin{figure}[ht]
  \centering
  \begin{minipage}{7in}
    \includegraphics[width=\linewidth]{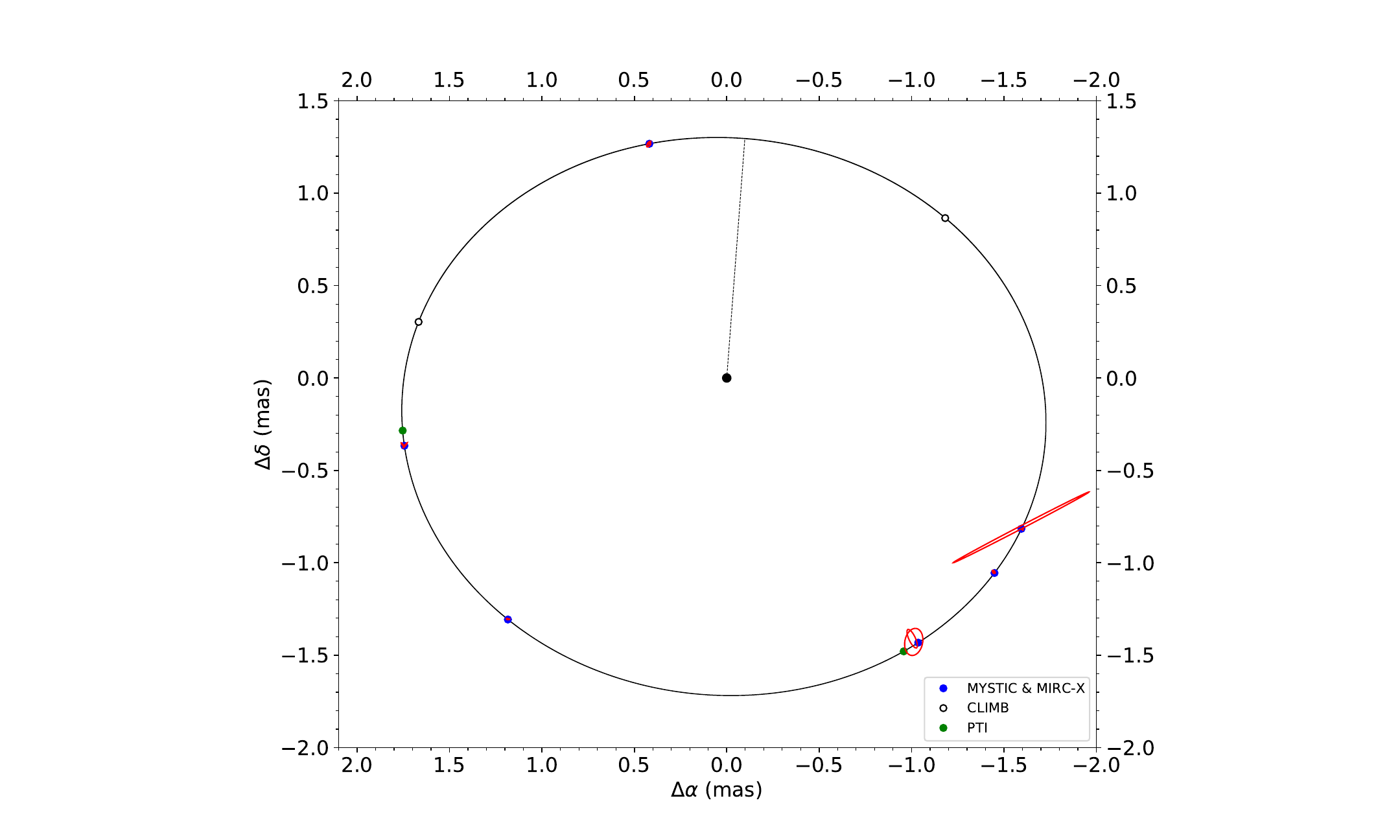}
    \caption{Sky plane orbit for HD 21278 B relative to HD 21278 A.  The white circles show the model positions at the time of the CLIMB observations, while blue points are model positions at the time of MIRC-X and MYSTIC observations.  Green points show model positions for PTI.  The red ellipses around the MIRC-X points are centered on the measured positions, and indicate $1\sigma$ uncertainties.  Some of the uncertainty ellipses are smaller than the plotted model points.
  \label{fig3.4}}
  \end{minipage}
\end{figure}

Nightly relative positions were calculated using a binary grid search program\footnote{ \url{https://www.chara.gsu.edu/analysis-software/binary-grid-search}} \citep{Schae16}, fitting to visibility and closure phase measurements from multiple baselines. The position measurements from the CHARA array are given in Table \ref{tab3} below. The individual position measurements in Table \ref{tab3} were not used in our binary orbit fits, but are provided for illustration of the orbit constraints.  Column 2 gives the date of observation, column 3 gives the angular separation of the binary $\rho$, column 4 gives the position angle of the secondary star $\theta$ (measured counterclockwise from north), columns 5 and 6 give the major and minor axis sizes of the error ellipse respectively, and column 7 gives the orientation angle of the error ellipse $\varphi$. The errors in the binary positions were computed using a Monte Carlo bootstrap approach with 1000 iterations. During each iteration, measurements were randomly selected with repetition to construct a new sample of visibilities and closure phases that were randomly varied assuming Gaussian uncertainties. The uncertainties in the binary positions are based on the $67.5\%$ confidence ellipses for two parameters from the bootstrap distributions.
As shown in Fig.~\ref{fig3.4}, the sky orbit agrees well with the relative position measurements determined from individual nights of interferometric data.  

\begin{deluxetable}{cllrllrl}

\tabletypesize{\footnotesize}
\tablewidth{0pt}
\tablecaption{Nightly Position Fits for HD 21278 \label{tab3}}
\tablehead{
\colhead{UT Date} & \colhead{Time} & \colhead{$\rho$} & \colhead{$\theta$} & \colhead{$\sigma_{\rm maj}$} & \colhead{$\sigma_{\rm min}$} & \colhead{$\varphi$} & \colhead{Combiner} \\
\colhead{(MM/DD/YYYY)} & \colhead{(day)} & \colhead{(mas)} & \colhead{($^\circ$)} & \colhead{(mas)} & \colhead{(mas)} & \colhead{($^\circ$)} & \colhead{}
}
\startdata 
11/18/2021 & 59536.371 & 1.7840 & 233.992 & 0.0084 & 0.0050 &  13.958 & MIRC-X \\
11/18/2021 & 59536.371 & 1.7860 & 234.025 & 0.0107 & 0.0056 &  42.223 & MYSTIC \\
12/11/2021 & 59559.368 & 1.7312 & 215.424 & 0.0554 & 0.0175 & 25.264 & MIRC-X\\
12/11/2021 & 59559.368 & 1.7504 & 215.318 & 0.0743 & 0.0483 & 168.683 & MYSTIC\\
12/19/2021 & 59567.368 & 1.7909 & 101.627 & 0.0122 & 0.0028 &  36.264 & MIRC-X\\
12/19/2021 & 59567.368 & 1.7776 & 101.823 & 0.0175 & 0.0029 &  140.298 & MYSTIC\\
11/14/2022 & 59897.372 & 1.3398 & 18.300 & 0.0138 & 0.0050 & 133.734 & MIRC-X\\
11/14/2022 & 59897.372 & 1.3300 & 18.520 & 0.0143 & 0.0068 & 133.068 & MYSTIC\\
11/21/2022 & 59904.362 & 1.7987 & 242.929 & 0.0051 & 0.0028 & 4.514 & MIRC-X\\
11/21/2022 & 59904.362 & 1.7860 & 243.093 & 0.4143 & 0.0080 & 117.501 & MYSTIC\\ 
9/28/2023 & 60215.314 & 1.7632 & 137.716 & 0.0066 & 0.0034 & 74.976 & MIRC-X\\
9/28/2023 & 60215.314 & 1.7569 & 137.982 & 0.0029 & 0.0013 & 131.763 & MYSTIC\\
\enddata
\vspace{0.5cm}
\tablecomments{Time = HJD - 2,400,000}
\end{deluxetable}

\section{Analysis}

The characteristics of the HD 21278 system imply that the stars had some tidal interaction that slowed their rotation compared to other $\alpha$ Per B-type stars. This most likely happened during the pre-main sequence evolution. That said, there are not signs of strong present-day interactions that would have disrupted the evolution of the individual stars, and we will treat them as having evolved as single stars in the remainder of the paper. The near future will bring the expansion of the primary star and mass transfer to the secondary star before it begins stable core helium burning.

\subsection{Age Determination}

We produced a color-magnitude diagram (CMD) of the members of $\alpha$ Persei by matching 2MASS data \citep{Cut03,Skr06} 
with the \cite{Boy23} Gaia DR3 membership catalog.  2MASS photometry was used in our primary analysis because the interferometric luminosity ratios for the binary's components were determined at wavelengths within the 2MASS $H$ and $K_s$ bandpasses.  
\cite{Boy23} corrected the photometry in their study for extinction using STILISM dust maps \citep{Lal18,Cap17} to return individualized reddening $E(B-V)$ values based on sky position and distance.  

\begin{figure}[ht]
  \centering
  \begin{minipage}{4.5in}
    \includegraphics[width=\linewidth]{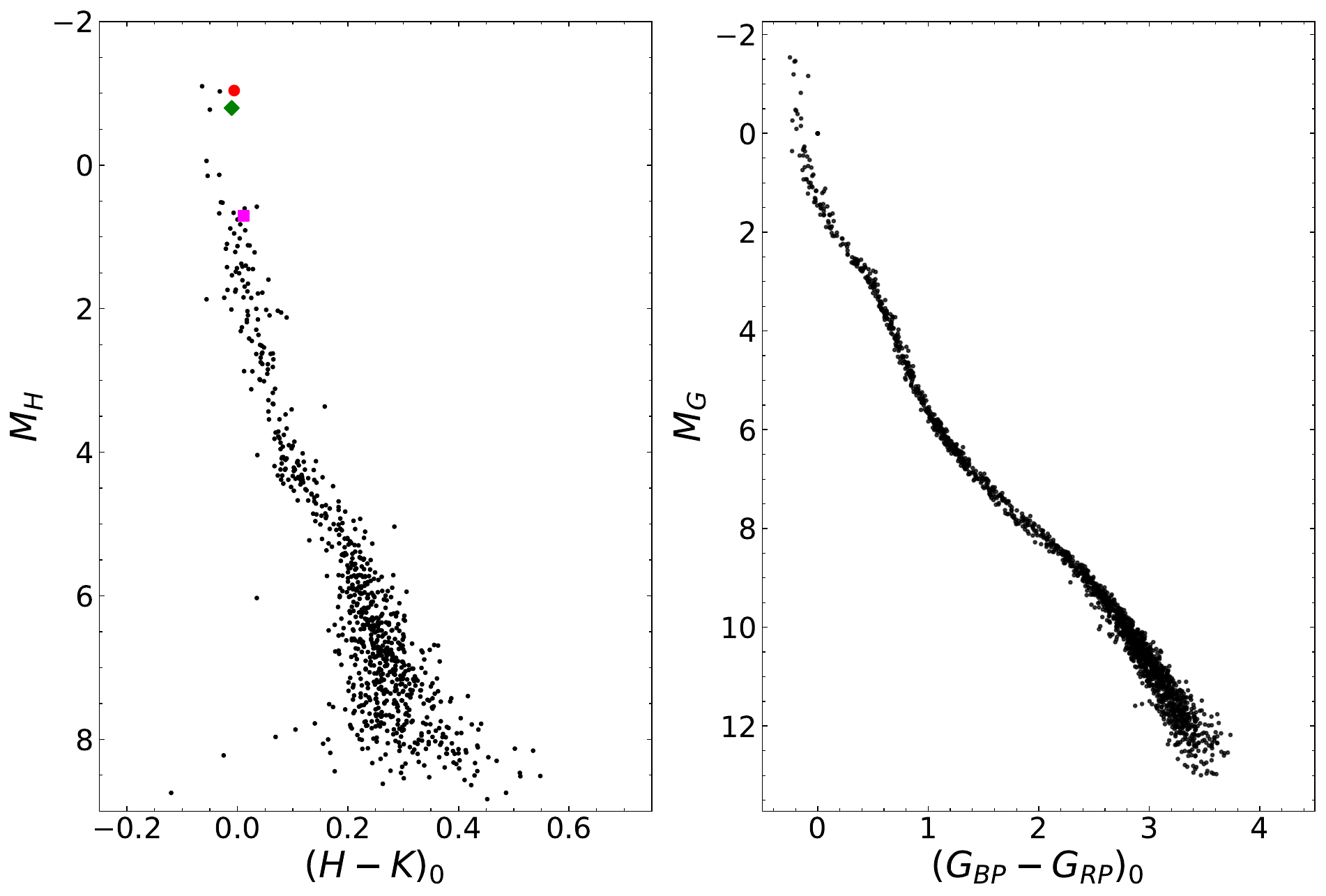}
    \caption{{\it Left:} CMD of the $\alpha$ Persei cluster using 2MASS photometry cross-referenced with \cite{Boy23} membership information. HD21278 (red), the primary star (green), and secondary star (magenta) are shown. {\it Right:} CMD of $\alpha$ Persei using Gaia DR3 photometry.
  \label{fig4.1}}
  \end{minipage}
\end{figure}

We used the STILISM-based reddening values to calculate extinctions in 2MASS $H$ and $K_s$ bands using the extinction law from \cite{Car89} for the majority of the 2MASS data.  We chose to apply the reddening value that we found using the SED fits discussed in the Appendix for HD 21278.  
Using Gaia parallax data from \cite{Boy23} and the extinctions, the $H$ and $K_s$ apparent magnitudes were converted to absolute magnitudes, as shown in the CMD in Fig.~\ref{fig4.1}. We also took the total absolute magnitude of the HD 21278 binary, and calculated the photometry of the two component stars using the measured flux ratios from Table \ref{tab2} and the distance measured from \cite{Bai21}.

The absolute magnitudes of the primary and secondary stars are affected by uncertainty in distance, apparent $H$ and $K_s$ magnitude, $H$ and $K_s$ extinction, and $H$ and $K_s$ flux ratio. We mapped out the likely color-absolute magnitude diagram (CAMD) positions of the two component stars using a Monte Carlo (MC) simulation, assuming the different uncertainties were uncorrelated (except for the extinctions). We generated $10^4$ points from Gaussian distributions for each source of uncertainty.  We contoured the density of the resulting values for both the primary and the secondary, and show the contour containing 68 percent of the points. 
The binary's age was determined by comparing the 2MASS CAMD position of the two stars with isochrones from PARSEC \citep{PARSEC} and MIST \citep{Cho16, Dot16, Pax11, Pax13, Pax15}, and are based primarily on the agreement between the isochrone predictions and the uncertainty regions for the two stars. 

\begin{figure}[ht]
  \centering
  \begin{minipage}{4.5in}
    \includegraphics[width=\linewidth]{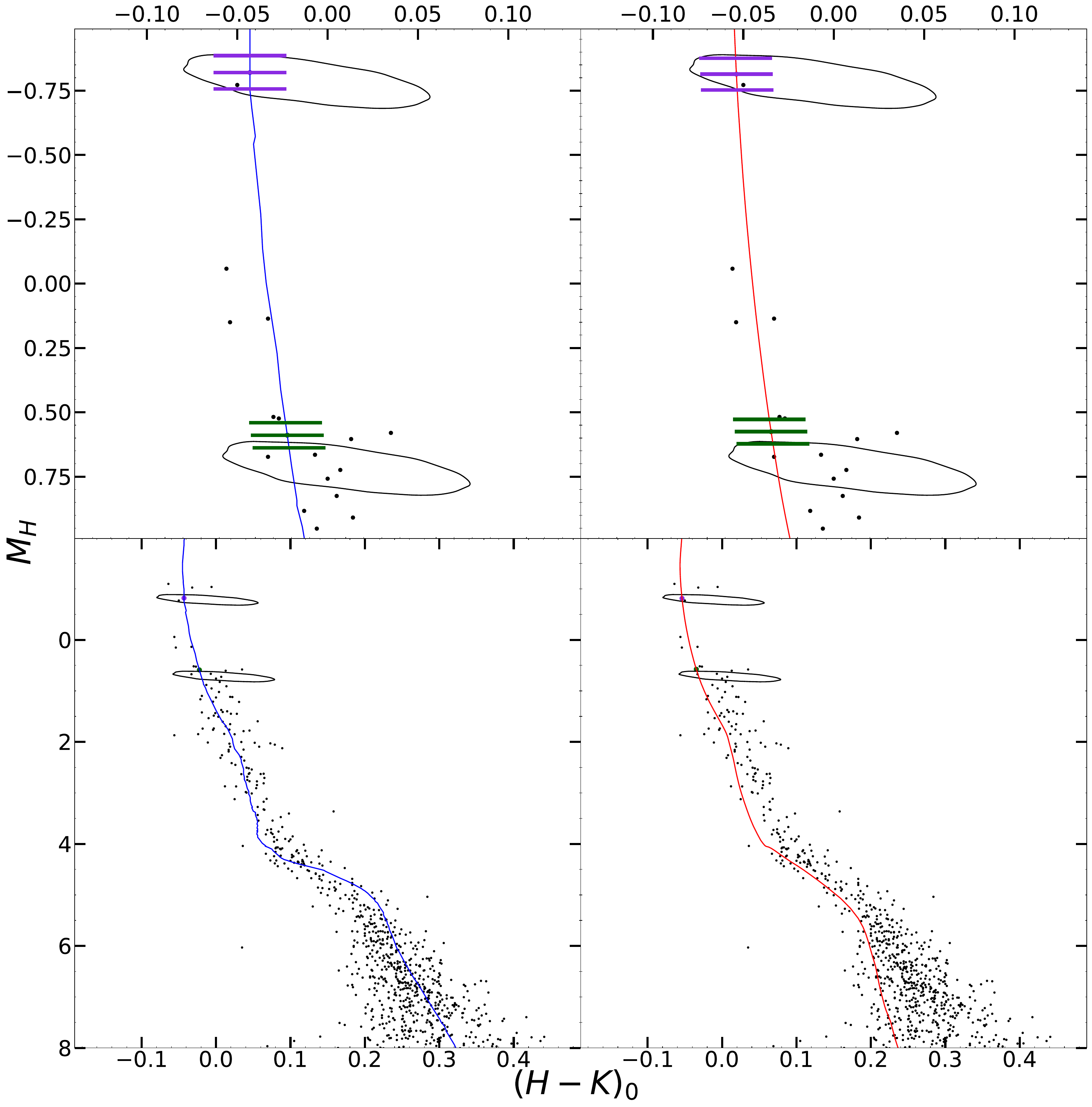}
    \caption{{\it Left:} Comparison of a PARSEC isochrone (age 49 Myr, $\omega$=0.2) with the 2MASS CAMD of $\alpha$ Persei. The uncertainty ellipses for the primary and secondary star positions are shown, along with isochrone predictions for their masses (and $1\sigma$ uncertainties). {\it Right:} Same, but for a MIST isochrone at an age of 49.5 Myr.  Since the current version of MIST cannot make a model with a rotation of $\omega= 0.2$, we used the best fitting model for $\omega=0.4$.
  \label{fig4.2}}
  \end{minipage}
\end{figure}

We used PARSEC version 2 in our primary comparisons.  Even though both stars in HD 21278 have fairly slow rotation rates for B-type stars, they are still rotating at a fast enough rate to have an effect on the age estimate, so we use $\omega = 0.2$ as most appropriate for the primary star. As shown in Fig.~\ref{fig4.2}, the PARSEC isochrones prefer an age of $49 \pm 7$ Myr. We can make some estimations of systematic errors through comparisons involving other isochrone sets. 
Using MIST isochrones, which have approximately the same amount of convective core overshooting, we found ages at $\omega = 0.0$ and $\omega = 0.4$ and interpolated.  This returned an age of $49.5 \pm 6$ Myr.

Because PARSEC v1.2S and v2 isochrones use different amounts of convective core overshooting (with v1.2S having a larger amount), an age comparison can be used to estimate systematic uncertainty due to this physics uncertainty. We also compared to PARSEC v2 isochrones at $\omega = 0$ in order to estimate uncertainty due to model parameterization.
We find the overshooting contribution to be about 1 Myr and the rotation contribution to be about 2 Myr, for a systematic uncertainty of 3 Myr.  For all isochrone models, the metallicity uncertainty listed in section 1.1 is taken into account.  In all cases, a $\pm 0.03$ change in [Fe/H] adds about 2 Myr to the age range. Since the reddening value that we used came from a different source compared to the rest of the reddening values for $\alpha$ Persei, we compared the effect it would have on the age estimates for both PARSEC and MIST.  In both cases, the reddening value that we found resulted in the age decreasing by about $1-2$ Myr and the age uncertainty decreasing by $\pm1$ Myr.

Although our age estimate for $\alpha$ Persei is significantly lower than most previous age estimates, it is consistent with the lower end of the range of previous estimates. 
\citet{MWSC} found an age of about $50\pm12$ Myr, using 2MASS photometry and Padova isochrones. \citet{david} used Stromgren photometry for cluster B and A stars, and found an age of approximately 50 Myr from isochrone fits in the $\teff- \log g$ plane. \citet{IACOB} discuss the spectroscopic classification of the cluster's B stars, and conclude that the presence of four B3V-type main sequence stars argues for an age in the $50-60$ Myr range. We discuss the characteristics of cluster's B stars in more detail in Appendix \ref{bseds}, and find additional evidence in support of this age.  
 By contrast, the lithium depletion boundary (LDB) age ($79^{+1.5}_{-2.3}$ Myr; \citealt{galindo}) is significantly larger. While this age technique is thought to be physically straightforward and reliable, it is frequently found to return greater ages than other techniques do \citet[e.g.,][]{Dah15}, which may be connected to magnetic activity and spotting among the low-mass pre-main sequence stars. Straight isochrone fitting to the upper main sequence in young cluster CMDs is often difficult due to small numbers of stars and age insensitivity of the main sequence shape. Precise masses and evolutionary information help us to largely avoid these problems.

\begin{figure} [ht]
  \centering
  \begin{minipage}{4.5in}
    \centering
    \includegraphics[width=0.5\linewidth]{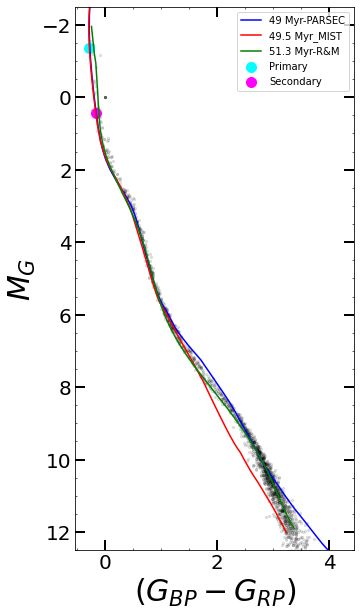}
    \caption{Comparison of best-fit isochrones and the $\alpha$ Per CMD using Gaia DR3 data from \cite{Boy23}.  The MIST isochrone (red) uses $\omega=0.4$, and an empirical isochrone \citep{Rot24} is also overlaid in green.
  \label{fig4.3}}
  \end{minipage}
\end{figure}

\subsection{Ultramassive White Dwarfs}
With a new age estimate for the cluster, we can now revisit initial mass estimates of massive white dwarfs that may have originated from $\alpha$ Persei. \cite{Mil22} identified three white dwarfs that may have started as members of the cluster, but eventually escaped.  These WDs were identified by taking the proper motions and positions of massive WDs ($M > 0.85 M_{\odot}$) in a Gaia EDR3 white dwarf catalog \citep{Gen21} and back-tracing their paths.  If at any point within the kinematic age estimate ($81\pm 6$ Myr; \citealt{Heyl21}), a WD candidate was within 15 parsecs of $\alpha$ Persei, it was treated as a possible member of the cluster.  \citeauthor{Mil22} found a total of 5 possible candidates, but two were removed from the sample because their estimated cooling times are much greater than the estimated age of the cluster.

For the remaining three candidates, \cite{Mil22} used WD models to constrain the final masses and cooling times.  None of the candidates have a strong magnetic field based on a lack of measurable Zeeman splitting, indicating that they are likely not the result of a merger. Table \ref{tab4} gives the WD masses (measured from Balmer line spectroscopy), estimated time since escape, and cooling times and initial masses inferred by \citet{Mil22} using the 81 Myr kinematic age with an assumption that the WD cores have an oxygen-neon (ONe) composition. Each of the progenitor masses were determined by using PARSEC isochrones \citep{Bre12,Tan14,Che14,Che15,Mar17,Pas19,Pas20}. 

We redetermined the initial masses of the WD candidates using our cluster age estimates and the \citet{Mil22} cooling times.  %In order to minimize systematic uncertainty, 
We used non-rotating PARSEC v.2 isochrones at the same metallicity we employed for our analysis of HD 21278 for the initial mass estimates.
Uncertainties in the metallicity lead to minor effects on the initial mass estimates of $0.02-0.04 M_{\odot}$.  Our initial mass estimates are included in Table \ref{tab4},  where the initial mass uncertainties were determined by adding uncertainties due to cooling time and total age in quadrature.

\begin{deluxetable}{crcccccc}

\tabletypesize{\footnotesize}
\tablewidth{0pt}

\tablecaption{White Dwarf Candidate Initial-Final Mass Values \label{tab4}}

\tablehead{& & \multicolumn{4}{c}{\citet{Mil22}} & \colhead{PARSEC}  & \colhead{MIST} \\ \colhead{ID} &
\colhead{Gaia DR3 ID} & \colhead{$M_{\rm final} / M_\odot$} & \colhead{$t_{\rm escape}$ (Myr)} & \colhead{$t_{\rm cool}$ (Myr)} & \colhead{$M_{\rm init} / M_\odot$} & \colhead{$M_{\rm init} / M_\odot$} & \colhead{$M_{\rm init} / M_\odot$}
}
\startdata 
WD1 & 439597809786357248 & $1.20\pm 0.01$ & 5 & $45\pm 4$ & $8.5\pm 0.9$ & $--$ & $--$ \\
WD2 & 244003693457188608 & $1.17\pm 0.01$ & 12 & $14\pm 4$ & $6.3\pm 0.3$ & $8.50 ^{+1.17}_{-0.85}$ & $8.38 ^{+0.96}_{-0.74}$ \\
WD5 & 1983126553936914816 & $1.12\pm 0.01$ & 30 & $3\pm 1$ & $5.9\pm 0.2$ & $7.45 ^{+0.62}_{-0.48}$ & $7.37 ^{+0.50}_{-0.41}$ \\
\enddata
%\vspace{-0.5cm}
\end{deluxetable}

Our age estimate for the cluster conflicts with the cooling time for the most massive white dwarf escapee candidate (WD 1, Gaia DR3 439597809786357248) from \cite{Mil22}. WD 1 has a cooling time of $45\pm4$ Myr which, when combined with the $51\pm7$ Myr age of the cluster, means that the progenitor of WD 1 would have had at max a $17$ Myr lifespan before becoming a white dwarf.
This probably indicates that WD 1 is not a former member of the cluster, despite its relative proximity. Potentially WD 1 might have been produced in a star formation complex that is related to $\alpha$ Per, but further attention to this WD seems warranted.  

\begin{figure}[ht]
  \centering
  \begin{minipage}{6in}
    \includegraphics[width=\linewidth]{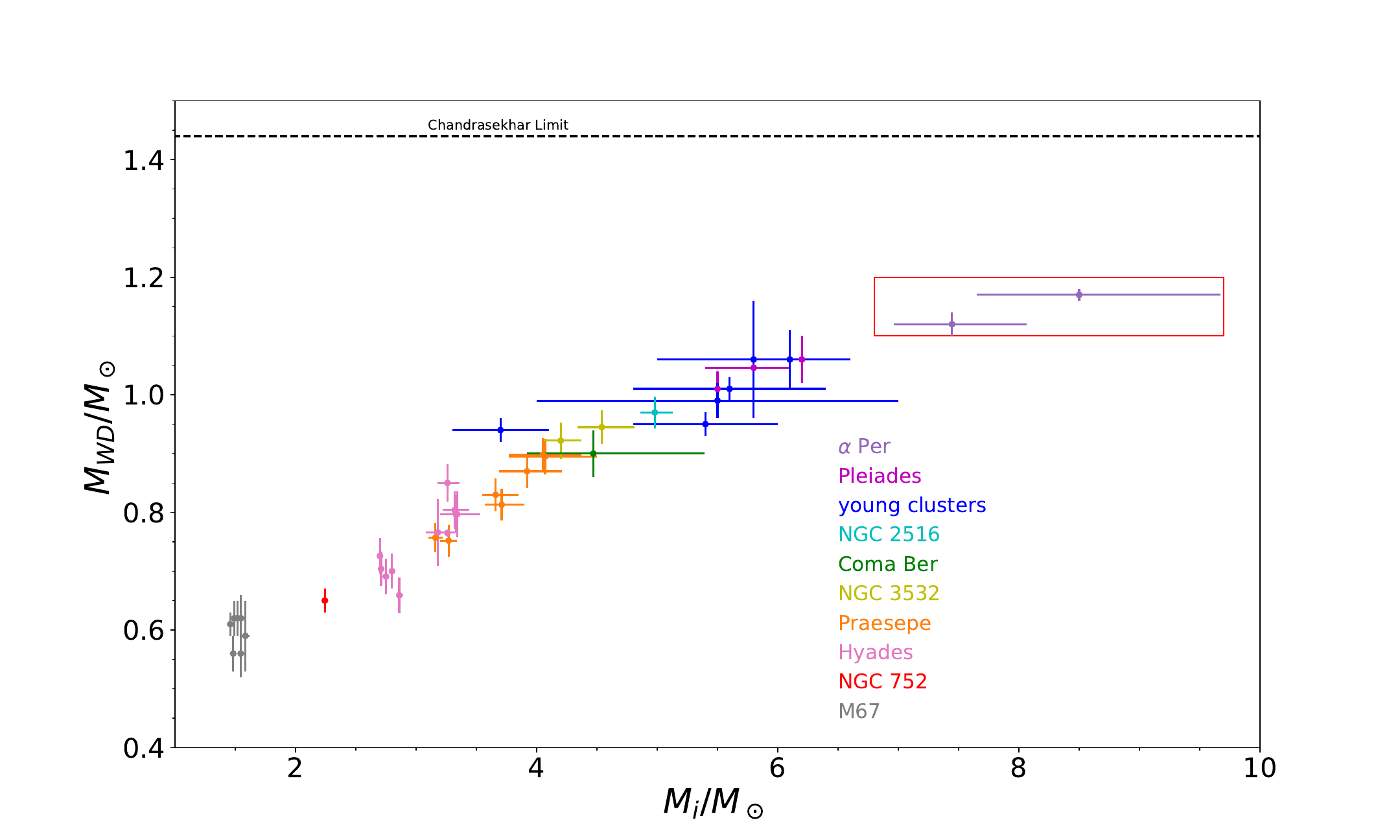}
    \caption{The initial-final mass relation for white dwarfs from measurements in open clusters. $\alpha$ Persei measurements are from this work and have been highlighted via the red box. Sources for data on other cluster WDs are: Pleiades, \cite{Hey22}; "young clusters", \cite{Ric21}; NGC 752, \cite{Mar20}; M67, \cite{Can21}; the Hyades, NGC 2516, and NGC 3532, \cite{Cum18}; Coma Ber, WD data from \cite{Dob09}, initial mass \cite{Lam23}; and Praesepe, WD data \cite{Cum18}, initial masses \cite{Mor22}.
  \label{fig5.1}}
  \end{minipage}
\end{figure}

The other two WDs remain viable candidates for former $\alpha$ Per members, although WD 5 is at a relatively large distance from the cluster center and would have had to have escaped while it was still a main sequence star. We plot the two WDs in the initial-final mass relation (IFMR) for open cluster WDs under the assumption that they were born with the rest of $\alpha$ Per. In Fig.~\ref{fig5.1}, we plot other initial mass values from the literature that were obtained using PARSEC isochrones in order to avoid systematic uncertainties deriving from different code physics. 
The revised initial masses for the $\alpha$ Per escapees plotted in Fig. ~\ref{fig5.1} indicate a nearly linear trend in the white dwarf IFMR at the high-mass end that continues into the range inferred for the minimum mass for supernovae \citep{smartt}. The evidence is an indication that white dwarfs generated from stars in the $7-8.5 \msun$ range may not reach the Chandrasekhar limit, or that the behavior of stars near the traditional minimum mass for supernova explosions may be more complicated than expected.

There are precious few observational probes of the high-mass end of the IFMR. In a piecewise-linear fit to the IFMR as constrained by Gaia field WDs within 100 pc, \citet{ElB18} find that their highest initial mass point ($8 \msun$) is consistent with Chandrasekhar mass, but weakly constrained ($1.37^{0.06}_{-0.21} \msun$). \citet{Cun24} conducted a similar study using a 40 pc sample of spectroscopically-characterized WD, but because their sample of massive white dwarfs was very small and did not have known progenitor masses, they calibrated the high-mass end of the IFMR against the high-mass end of an open and globular cluster WD sample from \citet{Cum18}. The most massive WDs in that sample (which are still significantly lower than the Chandrasekhar mass) are in the Pleiades and the AB Dor moving group, both significantly older than $\alpha$ Per. These candidate $\alpha$ Per WD candidates are therefore very valuable. It remains the case that these white dwarfs are no more than about $1.2 \msun$.

Before closing the discussion, we must point out that the unknown rotation rate in the progenitor stars of these WDs is a significant source of uncertainty.  Rapid rotation is expected to relieve some of the pressure on the core of a star, allowing nuclear evolution to proceed at a slower pace, leading to longer lifetimes. 
As discussed in the appendix, rapid rotation is common in the cluster's B stars, with some likely rotating close to critical. For HD 21278 A, however, binary interactions are likely to have kept the evolution fairly close to that of a non-rotating star. As a result, its age is likely to be more determinable than most other cluster stars. Based on the theory, it should be expected that differences in rotation from star to star will produce scatter in the IFMR, and that there will be a bias toward producing WDs from slower rotating stars earlier. For these $\alpha$ Per candidate WDs, the determination of the initial masses is complicated by our lack of information about how fast the progenitor stars were rotating. Because of this, our calculations using $\omega = 0$ isochrones primarily give us a lower limit for the initial masses of these WD candidates --- substantially faster rotation will push the initial masses well into an overlap with inferred minimum masses of supernova progenitors.  
Additional consequences of rotation on evolution and evolution timescales will need to be examined to clarify (if possible) the IFMR at the high-mass end.

\section{Conclusion}

Using a combination of spectroscopic and interferometric data, we fitted for orbital parameters of the HD 21278 binary.  Based on the best-fit model, the masses of the stars are $5.348\pm 0.085 \msun$ and $3.331\pm 0.062 \msun$.   Using the masses and the infrared photometric properties of the component stars, we find ages of $51 \pm 7$ Myr from PARSEC isochrones, and $51 \pm 6$ Myr from MIST isochrones. The largest source of uncertainty in the current stellar masses is the radial velocity measurements of the two stars in HD 21278.  The precision of future age estimates of $\alpha$ Persei can be further improved with additional precise radial velocity measurements.

Using the PARSEC age estimate, we revisited initial mass estimates for three ultramassive white dwarfs that are thought to have originated from $\alpha$ Persei.  The most massive candidate ($M = 1.20 \msun$) likely did not originate from $\alpha$ Persei because its cooling time would have given the progenitor star only $13$ Myr before it left the main sequence and implying a progenitor mass of over $14 \msun$.  
The two remaining candidates were estimated to have initial masses with a lower limit of $8.27 ^{+1.04}_{-0.78} \msun$ and $7.30 ^{+0.57}_{-0.45} \msun$ using the IFMR. If the white dwarfs truly originated from the cluster, they imply that the most massive single stars that do not supernova may not produce Chandrasekhar-mass white dwarfs unless there is large change in the slope of the IFMR at the highest masses.

\begin{acknowledgments}
We gratefully acknowledge support from the National Science Foundation under grant AAG 1817217 to E.L.S.  This research made use of observations from the SIMBAD database, operated at CDS, Strasbourg, France; the WEBDA database, operated at the Institute for Astronomy of the University of Vienna; and spectral data retrieved from the ELODIE archive at Observatoire de Haute-Provence (OHP, http://atlas.obs-hp.fr/elodie/). 
This work is based upon observations obtained with the Georgia State University Center for High Angular Resolution Astronomy (CHARA) Array at Mount Wilson Observatory.  The CHARA Array is supported by the National Science Foundation under Grant No. AST-1636624 and AST-2034336.  Institutional support has been provided from the GSU College of Arts and Sciences and the GSU Office of the Vice President for Research and Economic Development.  MIRC-X and MYSTIC received funding from the European Research Council (ERC) under the European Union's Horizon 2020 research and innovation programme (Grant No. 639889).  We thank the other members of the MIRC-X and MYSTIC development teams for their contribution.
We thank N. Morrell and H. Abt for their ability and willingness to provide electronic copies of their KPNO spectra (and  D. Willmarth for facilitating the communication).
The spectroscopy observations made with the Nordic Optical Telescope were taken by a variety of observers over the course of two months.  
The Palomar Testbed Interferometer was operated by the NASA Exoplanet Science Institute and the PTI collaboration. It was developed by the Jet Propulsion Laboratory, California Institute of Technology with funding provided from the National Aeronautics and Space Administration. This work has made use of services produced by the NASA Exoplanet Science Institute at the California Institute of Technology. This research has used data from the AAVSO Photometric All-Sky Survey (APASS), funded by the Robert Martin Ayers Sciences Fund and NSF AST-1412587.  
S.K. acknowledges support from an ERC Consolidator Grant (Grant Agreement ID 101003096) and STFC Small Award (ST/Y002695/1).
\end{acknowledgments}

\bibliography{sample631}{}
\bibliographystyle{aasjournal}

\facilities{PO:PTI, CHARA (MIRC-X), CHARA (MYSTIC), KPNO:CFT, NOT (FIES), Mercator1.2m (HERMES), TBL (Narval)}

\appendix

\section{Spectral Energy Distributions (SEDs)\label{bseds}}

Thanks to the large body of photometry and flux-calibrated spectroscopy for stars in $\alpha$ Per, well-sampled spectral energy distributions can be compiled. Model fits can determine effective temperature $T_{\rm eff}$ and angular diameter $\theta$ (or alternately, bolometric flux $f_{\rm bol}$, which is related via $f_{\rm bol} = \theta^2\sigma T_{\rm eff}^4 / 4$). When there are good measurements in the ultraviolet, the reddening can be determined as well, with the assumption of an extinction law. This kind of analysis has been done previously by \citet{fm05} and \citet{gordonb}, primarily for field B stars.

We developed code to simultaneously fit photometry, spectroscopy, and spectrophotometry for the cluster B stars. The main pieces of this use the package {\tt pysynphot} for determining observed photometric fluxes from models, and the package {\tt emcee} for fitting and determining uncertainties. We interpolated between three solar-metallicity ATLAS9 models \citep{atlas9} with different $T_{\rm eff}$ at the same surface gravity to generate intermediate temperature models. We used the extinction law of \citet{gordon23} in fitting dust effects on the SED.

We fitted all of the known single B-type members of the cluster or its corona (see Table \ref{sedtab}), as well as binaries where the secondary star probably contributes relatively little to the total flux. We summarize the photometry and spectrophotometry sources and their flux calibrations in Table \ref{sedsources}. In addition, low-dispersion spectroscopy from IUE was available for most stars, and we applied \citet{mf2000} corrections to the spectra before combining multiple observations taken with the same camera (SWP, LWP, or LWR). We also use Hopkins Ultraviolet Telescope (HUT) spectra from the Final Archive \citep{hutfinal} for the Be stars $\psi$ Per and c Per, and a Wisconsin Ultraviolet Photo-Polarimeter Experiment (WUPPE) spectrum \citep{wuppe} for $\psi$ Per. (The WUPPE spectrum had to be scaled by a factor of 1.22 to match the HUT spectrum because of light losses from the small aperture.)

Uncertainties on the angular diameter are typically at the $1-2$\% level because $\theta$ is determined mainly by the flux normalization and there are many measures for each star. Uncertainties on $T_{\rm eff}$ were dominated mostly by uncertainty in the surface gravity --- changes of 0.25 in $\log g$ typically led to uncertainties in the range of $50-100$ K. Differences in the ultraviolet data available also lead to significant differences in $T_{\rm eff}$ uncertainty. From a systematic point of view, gravity darkening due to rapid rotation and uncertain inclination will tend to move stars toward lower measured temperature. For the measurement of radius, the distance is the largest contributor to the uncertainty quoted.

\begin{figure}[ht]
    \includegraphics[width=\linewidth]{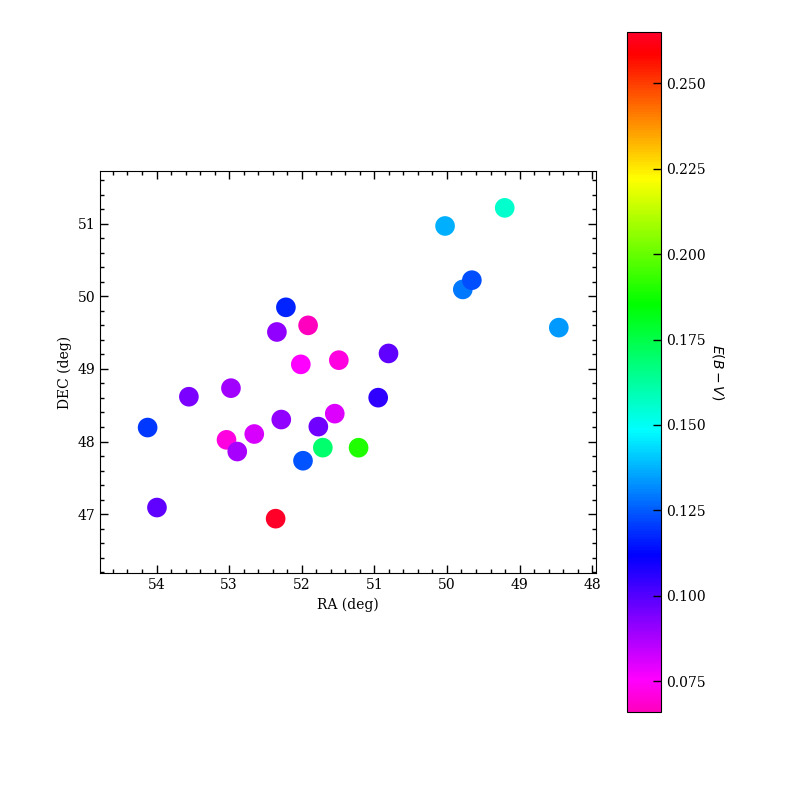}
    \caption{Measured reddening $E(B-V)$ for $\alpha$ Per B stars versus sky position.
  \label{aPermap}}
\end{figure}

\begin{figure}[ht]
    \includegraphics[width=\linewidth]{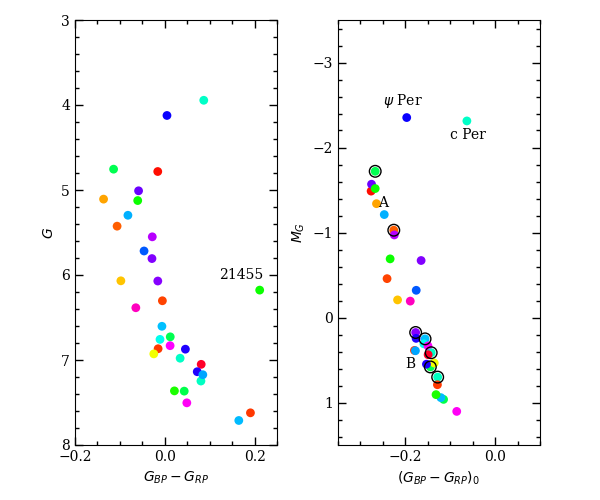}
    \caption{Color-apparent magnitude (left) and dereddened color -- absolute magnitude diagrams for $\alpha$ Per B stars. Color coding corresponds to projected rotation speed $v_{\rm rot} \sin i$. "A" is the inferred position of HD 21278 A after subtraction of the flux contribution of the secondary star proxy HD 21238 (labeled "B"). Known binaries with components that cannot be separated are circled in black.
  \label{aPcmds}}
\end{figure}

We find that there are substantial variations in reddening across the face of the cluster, as indicated in Fig. \ref{aPermap}. Similar variations have been identified by \citet{viln} using Vilnius photometric bands. While most of our reddening values hover around $E(B-V) = 0.10$, there appears to be an area of reduced reddening in the north part of the main body of the cluster (centered around HD 21278, 21071, 21091, and 21238). In addition, we identify enhanced reddening toward the star HD 21455, consistent with the detection of a molecular "cloudlet" by \citet{trapero}. To illustrate the fidelity of the reddening determinations, Fig. \ref{aPcmds} shows the {\it Gaia} color magnitude diagrams with and without distance and reddening corrections. The reduction in color scatter is immediately apparent, with the most significant effect on HD 21455.

\begin{deluxetable}{llcll}
\label{sedsources}
\tabletypesize{\scriptsize}
\tablewidth{0pc}
\tablecaption{Photometry and Spectrophotometry Sources and Flux Calibration for the SEDs}
\tablehead{
\colhead{Catalog} & \colhead{Filters} & \colhead{Wavelength} & \colhead{Photometry Ref.} & \colhead{Calibration Ref.}\\
& & \colhead{Range (\AA)} & & }
\startdata
TD1 & spectrophot. & $1380-2740$ & \citet{uvbssc} & \\
OAO2 & S4F4,S4F3,S4F1,S3F5,S3F1, & $1430-4255$ & \citet{code} & \citet{meade}\\
     & S2F1,S2F5,S3F2,S2F2,S1F4, & & \citet{meade} & \\
     & S1F1,S1F3 & & & \\
OAO2 & U2,U3 & $1621-2308$ & \citet{celescope} & \\
ANS & 15W,18,22,25,33 & $1550-3300$ & \citet{ans} & \citet{deboer}\\
TD1 & F1565,F1965,F2365,F2740 & $1565-2740$ & \citet{TD1} & \\
Gaia BP/RP Mean & spectrophot. & $3360-10200$ & \citet{gaianear} & \\
Johnson 13-Color & & $3374-11037$ & \citet{jandm} & \\
Vilnius & $UPXYZVS$ & $3450-6534$ & \citet{viln} & \\
& & & \citet{zandz} & \\
Geneva & $UB_1BB_2V_1VG$ & $3471-5814$ & \citet{geneva} & \\
Clampitt \& Burstein & & $3500-7400$ & \citet{clampitt} & \\
Str\"{o}mgren & $uvby$ & $3520-5480$ & \citet{paunzen} & \citet{gray}\\
& & & \citet{pena} & \\
SDSS & $ugr$ & $3551-6166$ & \citet{sdssbright} & \\
4-Color & $WBVR$ & $3554-7166$ & \citet{wbvrphoto} & \citet{mvb}\\
Homogeneous Means & $UBV$ & $3663-5448$ & \citet{mermubv} & \citet{bessell}\\
Photoelectric & $UBVRIJHKLMN$ & $3663-98704$ & \citet{mandm} & \\
Tycho & $BT,VT$ & $4220-5350$ & \citet{tycho2} & \citet{mvb}\\
APASS & $BVgri$ & $4361-7439$ & \citet{apass} & \\
PanSTARRS1 & $grizy$ & $4810-9620$ & \citet{ps1} & \\
Hipparcos & $Hp$ & 5176 & \citet{hippnew} & \citet{mvb}\\
Gaia EDR3 & $G_{BP},G,G_{RP}$ & $5051-7726$ & \citet{gaianear} & \\
TASS & $VI_C$ & $5448-7980$ & \citet{tass} & \\
2MASS & $JHK_{s}$ & $12350-21590$ & \citet{2mass} & \citet{cohen}\\
WISE & $W1,W2,W3,W4$ & $33526-220883$ & \citet{wise} & \citet{wise}\\
AKARI & $S9W,L18W$ & $82283-176094$ & \citet{akari_surv} & \\
\enddata

\tablerefs{OAO2: Orbiting Astronomical Observatory 2. ANS: Astronomical Netherlands Satellite. SDSS: Sloan Digital Sky Survey. 2MASS: Two-Micron All-Sky Survey. WISE: Wide-Field Infrared Survey Explorer.}
\end{deluxetable}

We discuss six stars in particular for their relevance to the binary or the most evolved cluster stars. 

\begin{figure}[ht]
    \includegraphics[width=\linewidth]{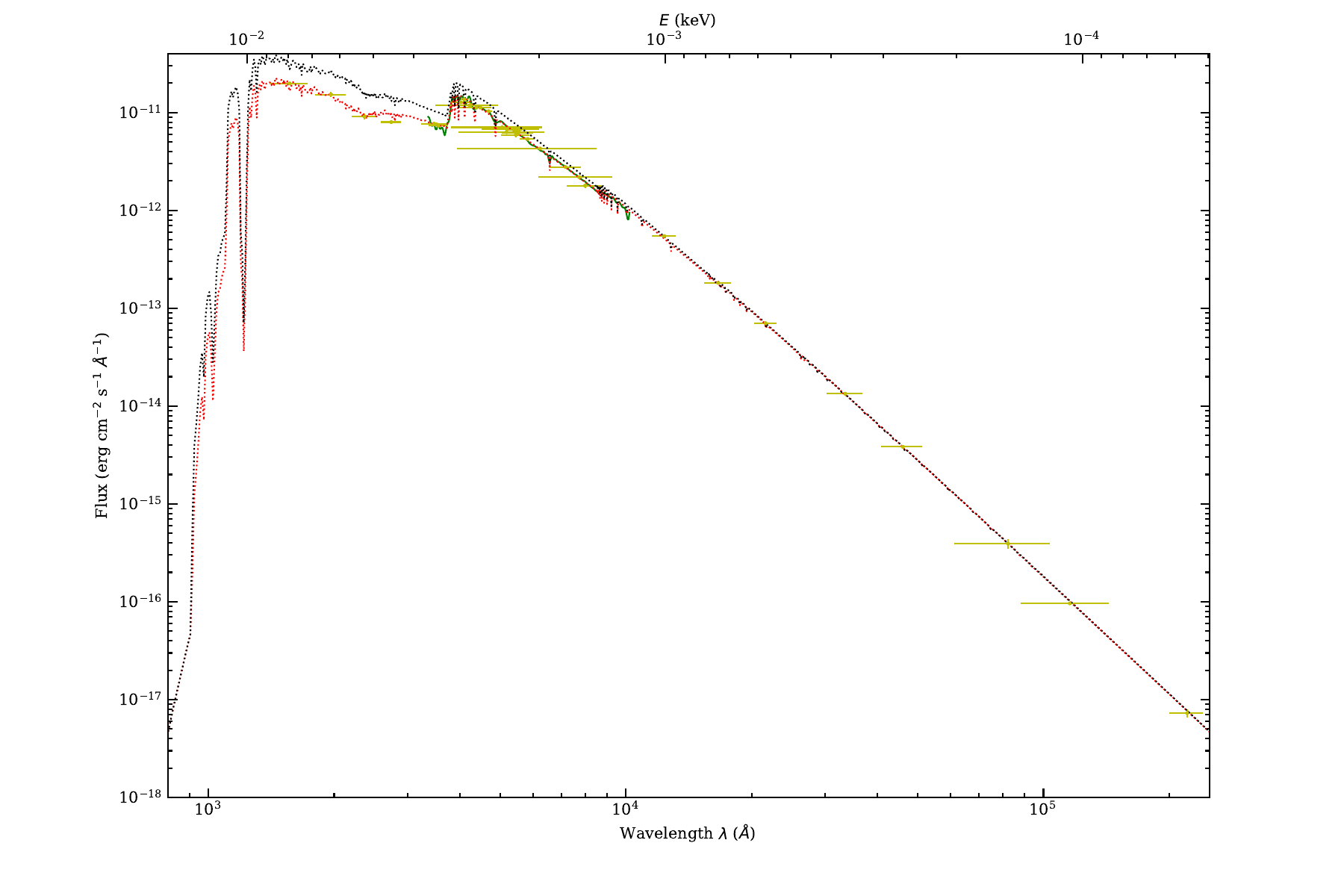}
    \caption{Spectral energy distribution for HD 21238. The green curve is the {\it Gaia} BP/RP spectrum, yellow points are photometric measurements, the black and red dotted lines are the unreddened and reddened ATLAS9 models, and red dots are the model predictions for the observed magnitudes.
  \label{hd21238}}
\end{figure}

{\it HD 21238:} This is the star whose infrared photometry most matches the inferred values for HD 21278 B from interferometric luminosity ratios. The only UV data available for $\lambda < 3000$ \AA \: comes from the TD1 satellite \citep{TD1}. The SED and fit are shown in Fig. \ref{hd21238} using models with $\log g = 4.25$.

{\it HD 21278:} We fitted the combined light for the stars in the binary, as well as an SED corrected for the light of the secondary using the proxy HD 21238. Because the secondary star contributes a significant amount of light in the optical and infrared ranges, a fit to the SED with the secondary star light subtracted should produce a higher $T_{\rm eff}$ and smaller $\theta$.

The model fit to HD 21238 was used to compute secondary star contributions in all photometric bands, even when HD 21238 was not observed in that band. The {\it Gaia} distances for HD 21238 and 21278 \citep{bailer} are consistent within the uncertainties, so that differential corrections for distance are small ($+0.024$ mag to account for HD 21238's smaller stated distance). Our fits also indicate that HD 21278 is just slightly more reddened than HD 21238, which indicates that the $(H-K_s)$ colors are close and also consistent within the measurement uncertainties. The two stars are relatively close together in the cluster, with lower reddening than the cluster average. 

We report the result with secondary star subtracted in Table \ref{sedtab}, which indicates that the primary star is probably one of the hottest of the main sequence stars, with a size consistent with an age of about 60 Myr. Because of the relatively slow rotation of the primary, we expect that the evolution of this star is close to what standard stellar models would predict.

\begin{figure}[ht]
    \includegraphics[width=\linewidth]{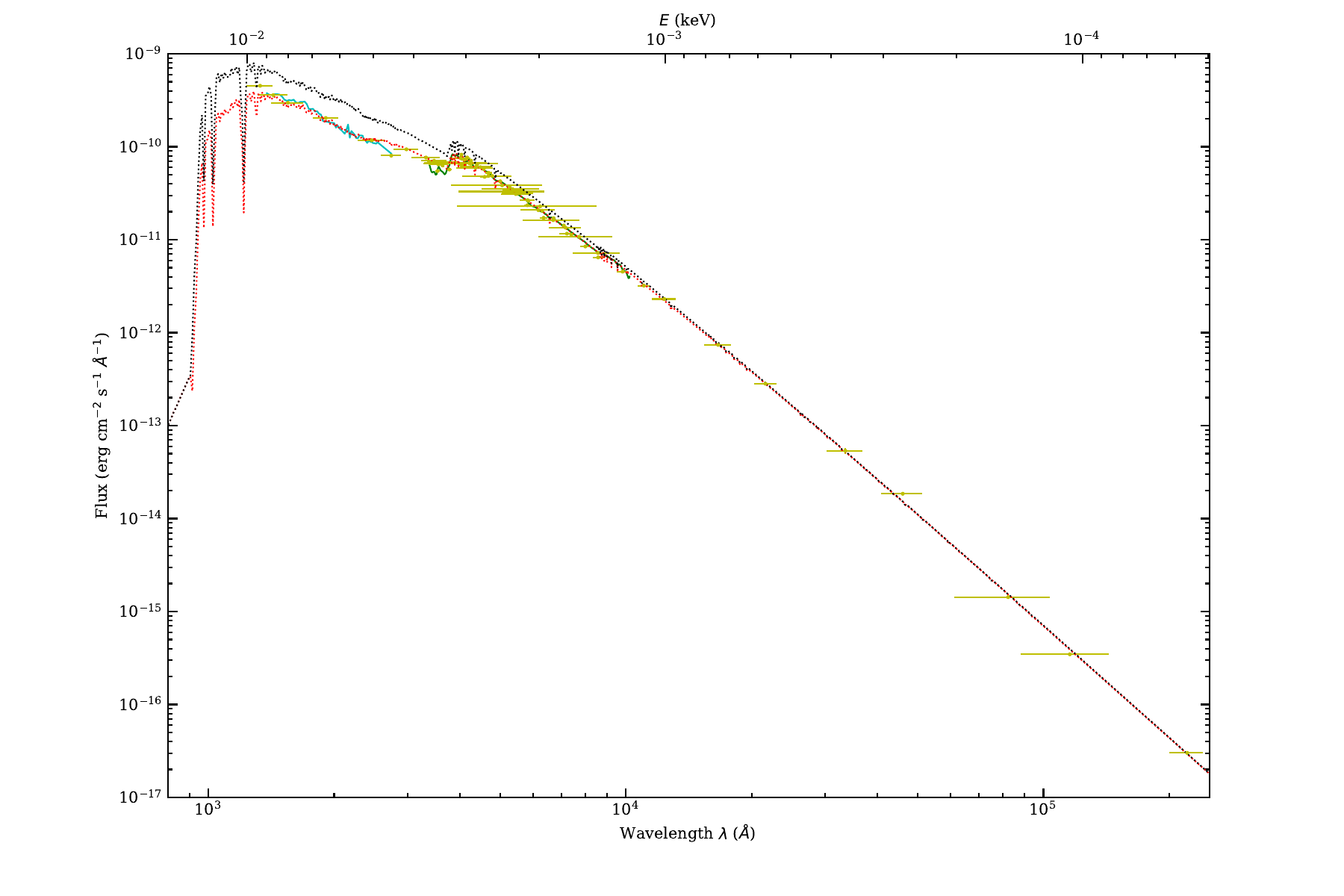}
    \caption{Spectral energy distribution for HD 21278 A. Points and lines have the same meaning as in Fig. \ref{hd21238}, but with the cyan line being ultraviolet spectrophotometry \citep{uvbssc}.
  \label{hd21278}}
\end{figure}

{\it Classical Be stars (c Per/48 Per/HD 25940, $\psi$ Per/37 Per/HD 22192, HD 21362, and HD 21455):} The luminosities of the brightest two stars (c Per and $\psi$ Per) identify them as the most massive and evolved B stars in the cluster. They are bright classical Be stars that are known to be rapidly rotating, and this is believed to be connected to disks detected in the infrared. Because this kind of disk is thought to be due to decretion from the star's equator resulting from rotation near breakup, the disk inclination is expected to mirror the star's spin inclination. An SED fit can give us information on the star's characteristics, but these will be affected by gravity darkening and need to be corrected to values that better represent what a non-rotating star would look like.

Interferometric measurements and H$\alpha$ emission line profiles tend to give similar inclination measurements for the disks. Using H$\alpha$ emission, \citet{delaa} find a double-peaked profile for $\psi$ Per, indicating a high inclination for the disk, and a single-peaked profile for c Per.
For c Per, the inclination is measured to be $45\pm5\degr$ \citep{jones17}, while $\psi$ Per is closer to equator-on at $i = 74\pm5\degr$ \citep{sg_be}. HD 21455 also has a fitted inclination $i=20\degr$ from the H$\alpha$ emission profile \citep{silaj}. While the measured rotation speeds $v_{\rm rot} \sin i$ for these stars are well below the expected critical rotation speed, this is expected to be affected by the spin inclination and by gravity darkening that de-emphasizes emitting regions at the equator. HD 21362 does not have an inclination measurement in the literature, but the observation of widely-separated H$\alpha$ emission peaks \citep{cotevank,ghosh} from the disk and the high $v_{\rm rot} \sin i$ measurement for the star imply a high inclination, which we estimate at $70\degr$.

For a rapidly rotating star, one of the challenges is finding measurable characteristics that would still be reflective of the star if it was not rotating. Polar radius is expected to differ by less than a couple of percent from that of a non-rotating star, and luminosity by less than 6\% for stars in the mass range relevant to $\alpha$ Per \citep{ekstrom08}. Even for rapid rotation, the distortion of the shape of the star affects a relatively small amount of mass in the surface layers, which means that conditions in the energy-generating core are only mildly affected. As for effective temperature, because rapid rotation makes the equator of the star expand and cool down while making the pole slightly contract inward and heat up, there will be a latitude on the star that retains the temperature of the non-rotating star. This is illustrated in Fig. 6 of \citet{ekstrom08}, which also indicates that a colatitude of $40^\circ$ retains the $\teff$ of the non-rotating star over nearly the whole range of $\omega = \Omega_{\rm rot} / \Omega_{\rm crit}$ values.

To model these Be stars, we employed FASTROT \citep{montesinos}, which uses a Roche model approximation to describe the surface, breaking the emission into contributions from patches that can have different $\teff$ and $\log g$. We take the spin inclinations of the stars as known, based on the observations of their disks, and only fit fluxes measured in wavelength ranges unaffected by the disk. (For c Per, we only used $\lambda < 7000$ \AA, while for the others we cut off at $10^4$ \AA.) We fit the SEDs to derive temperature at $40^\circ$ colatitude $T_{\rm eff,40}$, bolometric flux $f_{\rm bol}$, and reddening $E(B-V)$. We take the stellar luminosity from the model (as $f_{\rm bol}$ is dependent on viewing angle), and compute the representative radius for a non-rotating star from the luminosity and $T_{\rm eff,40}$. The results confirm the large radii for the two brighter stars, and for $\psi$ Per are consistent with combined star/disk SED modeling by \citet{klement17}: $R_p = 5.5\pm0.5 R_\odot$; $L = 2100\pm100 L_\odot$). ( To strict limits, \citealt{klement24} did not find a binary companion to $\psi$ Per, but found signs of a turndown in the SED of the disk that may indicate truncation by a faint companion. For 48 Per, a binary companion was identified, but contributing approximately 2\% of the light in $H$ band. We do not believe faint companions like these would significantly affect our conclusions on the characteristics of the star.)
There is an uncertainty in $T_{\rm eff,40}$ because the angular rotation rate $\omega$ is not well constrained by the SEDs, and so we assumed values between 0.8 and 0.9.

\begin{figure}[ht]
    \includegraphics[width=\linewidth]{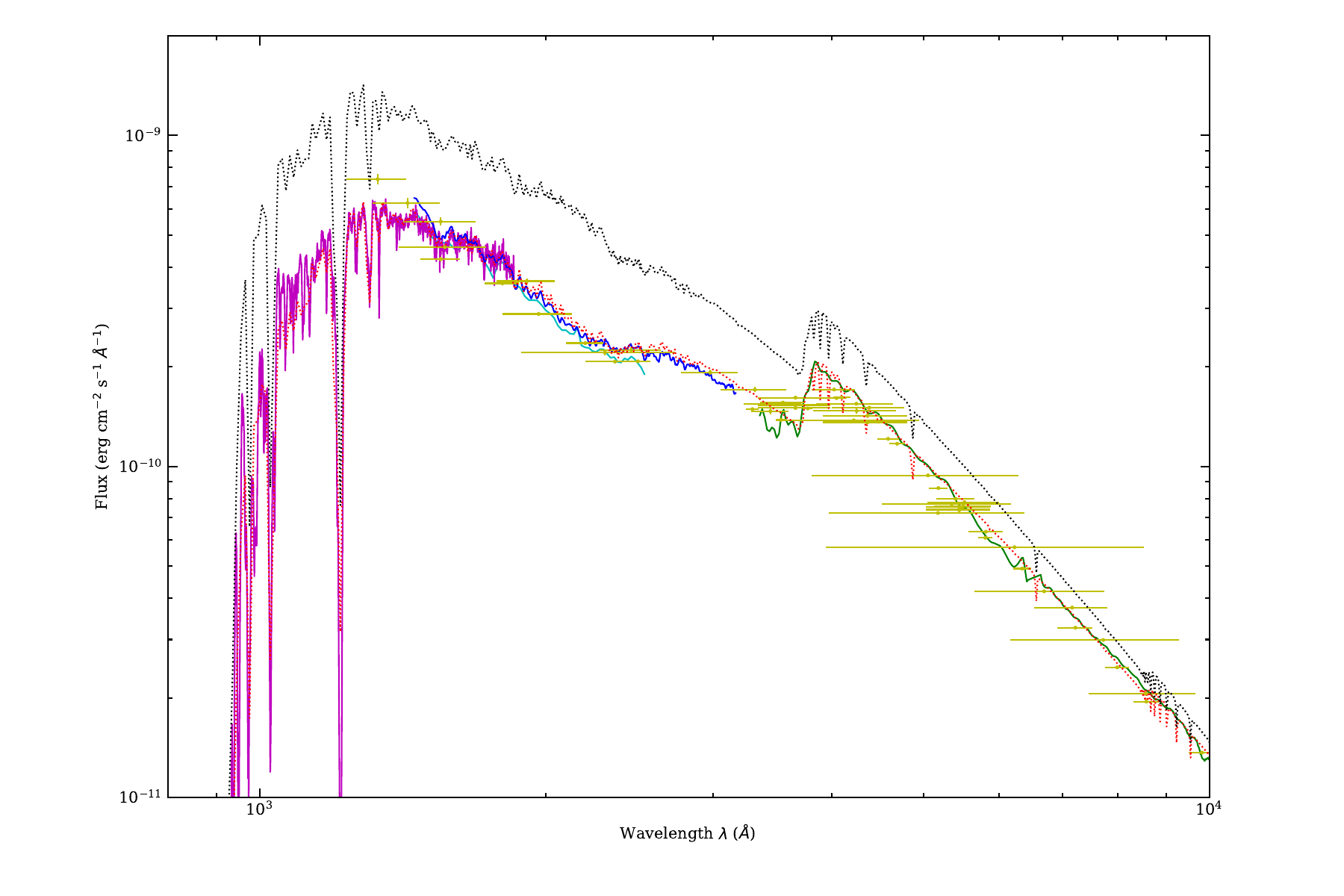}
    \caption{Spectral energy distribution for the Be star $\psi$ Per. Points and lines have the same meaning as in Fig. \ref{hd21238}, but with ultraviolet spectrophotometry \citep{uvbssc} in cyan, a HUT spectrum in magenta, and a WUPPE spectrum in blue (scaled to match the HUT spectrum in the overlapping wavelength range). The black and red dashed lines are the unreddened and reddened FASTROT model fits, respectively.
  \label{psiSED}}
\end{figure}

\begin{deluxetable}{llccccccl}[ht]
\label{sedtab}
\tabletypesize{\footnotesize}
\tablewidth{0pt}

\tablecaption{SED Fitting Results}

\tablehead{\colhead{HD} & \colhead{Alt. Name} & \colhead{Sp. Type} & \colhead{$T_{\rm eff}$} & \colhead{$\theta$} &\colhead{$E(B-V)$} & \colhead{$R / R_\odot$} & \colhead{$v_{\rm rot} \sin i$} & \colhead{Notes\tablenotemark{b}} \\ & & & \colhead{(K)} & \colhead{(mas)} & & & \colhead{(km s$^{-1}$)} &}

\startdata 
25490 & c Per / 48 Per & B5Ve & 15420\tablenotemark{a} & & 0.068 & 5.2 & 197 & classical Be\\
22192 & $\psi$ Per / 37 Per & B5Ve & 16690\tablenotemark{a} & & 0.087 & 5.2 & 275 & classical Be\\%Chauville+ 2001 vrot
\hline
21428 & 34 Per & B3V & 17720 & 0.243 & 0.091 & $4.55\pm0.20$ & 162 & VB\\
27396 & d Per / 53 Per & B4IV & 16270 & 0.270 & 0.155 & $4.19\pm0.08$ & 15 & SPB; corona\\
20418 & 31 Per & B3Vn & 16490 & 0.233 & 0.129 & $4.32\pm0.07$ & 300 \\
20365 & 29 Per & B3V & 17160 & 0.214 & 0.123 & $4.12\pm0.07$ & 139 \\ 
\hline
21278A & & B3V & 16410 & 0.205 & 0.076 & $3.75\pm0.09$ & 53\\
\hline
20809 & & B5V & 16080 & 0.201 & 0.098 & $3.77\pm0.06$ & 228 \\
21699 & & B9IIIp & 15100 & 0.191 & 0.071 & $3.63\pm0.09$ & 48 & He weak\\
21362 & & B6Vne & 15690\tablenotemark{a} &  & 0.078 & 3.2 & 319 & classical Be\\
27777 & 55 Per & B8V & 12600 & 0.199 & 0.077 & $3.10\pm0.05$ & 250 & corona\\
21551 & & B8V & 12590 & 0.191 & 0.081 & $3.62\pm0.10$ & 305 & Be \\
23383 & & B9Vnn & 11360 & 0.190 & 0.096 & $2.70\pm0.07$ & 306 & EB ($P=3.787$ d)\\
21071 & & B7V & 15020 & 0.143 & 0.071 & $2.51\pm0.02$ & 61 \\
21455 & & B7Vne & 13820\tablenotemark{a} &  & 0.234 & 2.5 & 139 & classical Be\\ 
19268 & & B5V & 15130 & 0.141 & 0.140 & $2.81\pm0.02$ & 28 \\
22402 & & B8Vn & 12980 & 0.141 & 0.074 & $2.83\pm0.02$ & 354 \\ 
21672 & & B8V & 12790 & 0.132 & 0.089 & $2.34\pm0.03$ & 223 & VB; SB1?\\
18538 & & B9V & 11790 & 0.135 & 0.093 & $2.17\pm0.01$ & 163 & SB\\
21641 & & B8.5V & 12400 & 0.127 & 0.088 & $2.35\pm0.02$ & 202 & Be; SB?\\ 
21181 & & B8.5V & 11670 & 0.131 & 0.096 & $2.46\pm0.02$ & 337 \\
22136 & & B8V & 13010 & 0.117 & 0.098 & $2.16\pm0.01$ & 25 \\ 
19624 & & B5 & 12140 & 0.130 & 0.132 & $2.46\pm0.01$ & 280 \\ 
\hline
21238 & & B9V & 11580 & 0.121 & 0.066 & $2.26\pm0.02$ & 79 \\
\hline
20863 & & B9V & 12090 & 0.120 & 0.105 & $2.30\pm0.02$ & 193 & VB\\ 
20510 & & B9V & 11540 & 0.127 & 0.137 & $2.36\pm0.02$ &\\ 
19893 & & B9V & 11290 & 0.123 & 0.134 & $2.28\pm0.01$ & 280 \\ 
20191 & & B9V & 12190 & 0.117 & 0.156 & $2.29\pm0.02$ & 230 \\
21279 & & B8.5V & 11860 & 0.111 & 0.124 & $2.05\pm0.01$ & 192 & SPB,SB,EB\\
21398 & & B9V & 11520 & 0.103 & 0.091 & $1.91\pm0.01$ & 135 & Be\\ 
21931 & & B9V & 11330 & 0.106 & 0.094 & $1.90\pm0.02$ & 161 \\ 
21091 & & B9.5IVnn & 10490 & 0.105 & 0.080 & $1.93\pm0.01$ & 340 \\ 
20961 & & B9.5V & 10990 & 0.110 & 0.190 & $2.10\pm0.01$ & 25 \\ 
21152 & & B9V & 10750 & 0.105 & 0.170 & $2.01\pm0.01$ & 225 \\ 
\enddata
\tablenotetext{a}{FASTROT fit to $T_{\rm eff,40}, f_{\rm bol}, E(B-V)$}
\tablenotetext{b}{"VB": visual binary. "EB": eclipsing binary. "SPB": slowly-pulsating B star. "corona": likely member of the cluster corona. "He weak": chemically peculiar star. "SB": spectroscopic binary.}
\end{deluxetable}

\begin{figure}[ht]
    \includegraphics[width=\linewidth]{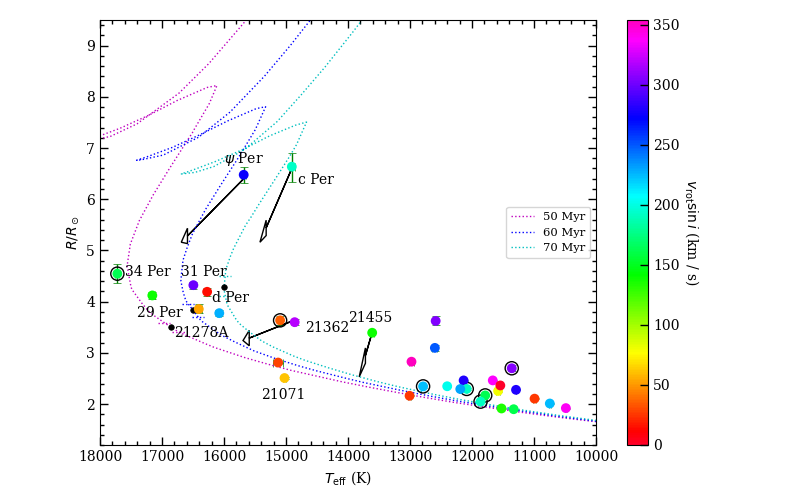}
    \caption{Radius versus temperature for B-type stars in $\alpha$ Per. Color coding corresponds to projected rotation speed $v_{\rm rot} \sin i$. For the Be stars, the fit using a single temperature photosphere is connected with the characteristics inferred from rotating models. Known binaries with components that cannot be separated are circled in black. PARSEC isochrones with $\omega=0.2$ for three ages are shown, with a black dot showing the location of a star with mass equal to HD 21278A.
  \label{aPrt}}
\end{figure}

The results for B stars in $\alpha$ Per are shown in Fig. \ref{aPrt}.  For the comparison of HD 21278A with PARSEC models having the proper rotation rate ($\omega = 0.2$), an age of about 61 Myr is implied. Rotation affects the positions of other stars in the figure in ways that are difficult to correct, but generally toward lower measured temperature and larger radius. The lower envelope is consistent with an age of 50 Myr.

\section{MYSTIC-Only Figures}

\begin{figure}[ht]
  \centering
  \begin{minipage}{6in}
    \includegraphics[width=\linewidth]{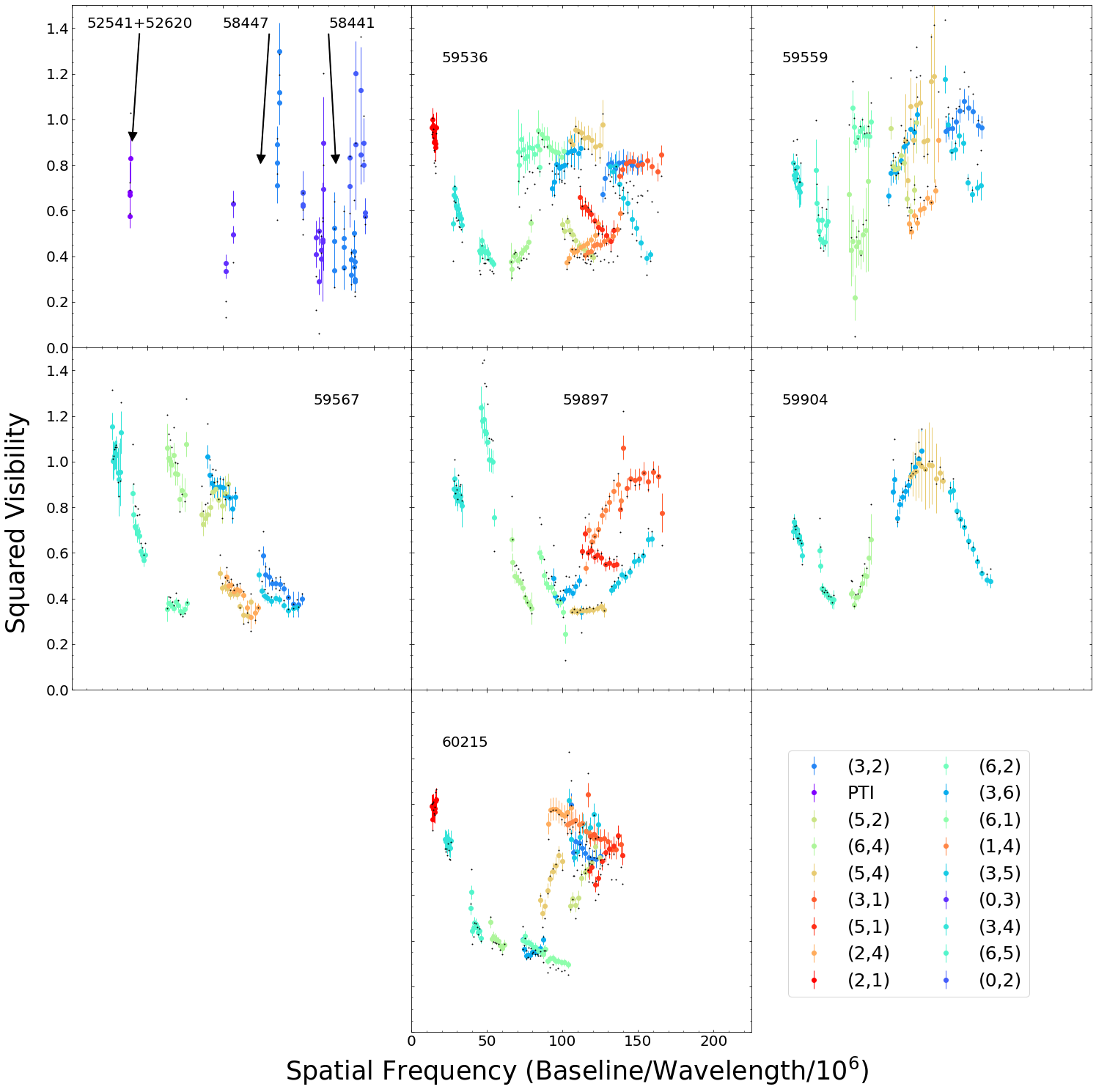}
    \caption{Squared visibilities versus spatial frequency for MYSTIC observations and predictions of the best-fit model (black dots), separated by  mJD of observation.  To reduce clutter, only observations taken at one epoch were used for each night.  Observations are separated by telescope pair with each pair denoting the two telescopes that were used to obtain squared visibility measurements. The top left panel has four epochs of PTI and CLIMB observations. 
  \label{append1}}
  \end{minipage}
\end{figure}

\begin{figure}[ht]
  \centering
  \begin{minipage}{6in}
    \includegraphics[width=\linewidth]{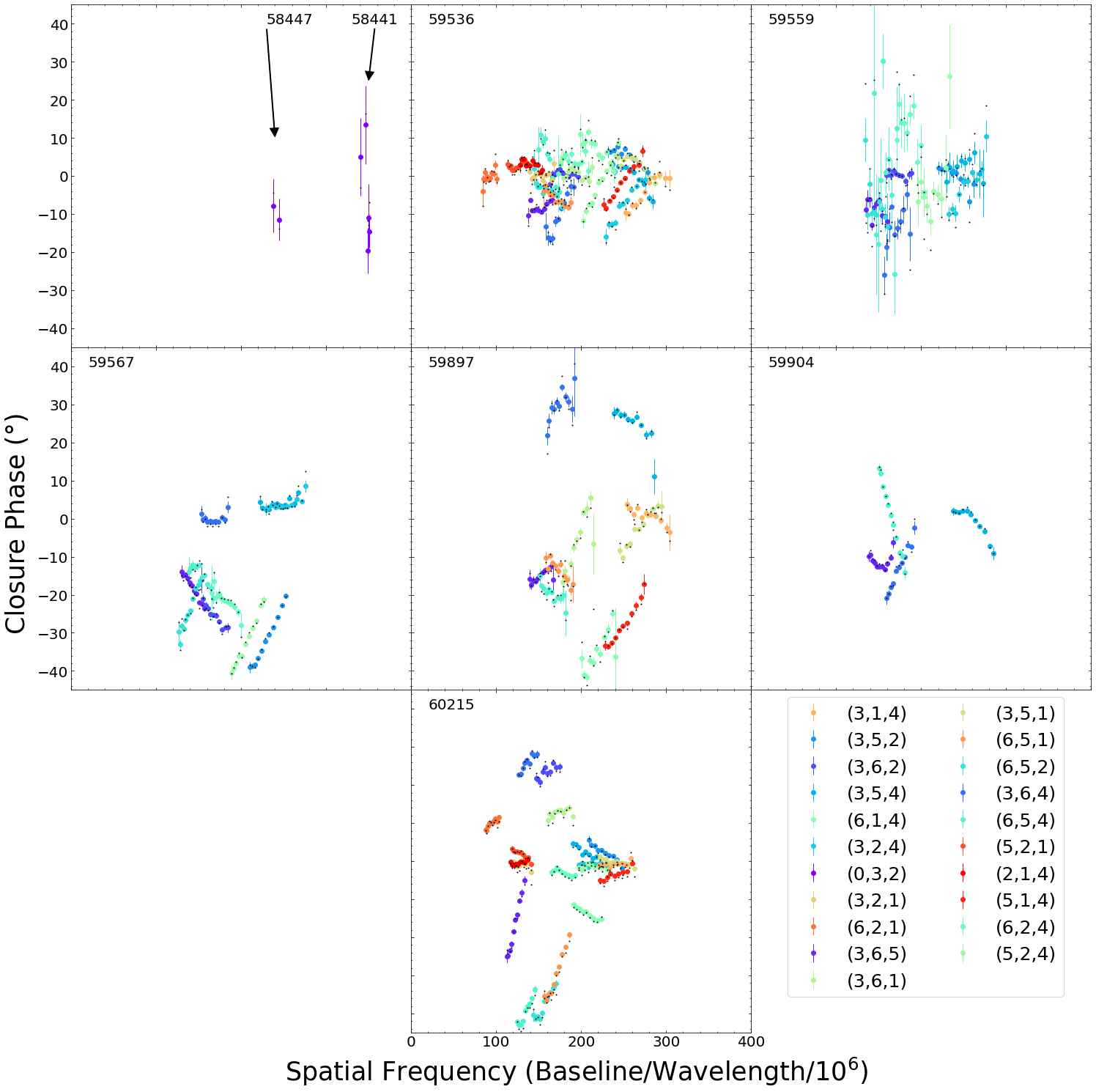}
    \caption{Closure phase versus spatial frequency for MYSTIC observations and predictions of the best-fit model (black dots), separated by mJD of observation.  The spatial frequency is obtained by taking the sum of the two baselines and dividing by the wavelength data.  To reduce clutter, only observations taken at one epoch were used for each night.  The top left panel has two epochs of CLIMB observations.  Each closure phase point is taken from 3 different telescope apertures, as listed in the legend.  
  \label{append2}}
  \end{minipage}
\end{figure}

\end{document}